\documentclass[a4paper,fleqn,usenatbib]{mnras}

\usepackage{newtxtext,newtxmath}
\usepackage[T1]{fontenc}
\usepackage{ae,aecompl}
\usepackage{dsfont}
\usepackage{color}
\usepackage{microtype}

\usepackage{cite}
\usepackage{rotating, tabularx}
\usepackage{bm}

\usepackage{graphicx}	% Including figure files
\usepackage{amsmath}	% Advanced maths commands
\usepackage{amssymb}	% Extra maths symbols

\usepackage[normalem]{ulem}   % for strikeout -- remove when sending back to MNRAS

\newcommand{\blobby}{\textsc{Blobby3D}}
\newcommand{\bbarolo}{\textsc{$^\mathrm{3D}$Barolo}}
\newcommand{\dnest}{\textsc{DNest4}}
\newcommand{\galpak}{\textsc{GalPak3D}}
\newcommand{\gbkfit}{\textsc{GBKFit}}
\newcommand{\tirific}{\textsc{TiRiFiC}}
\newcommand{\lzifu}{\textsc{LZIFU}}
\newcommand{\kinemetry}{\textsc{Kinemetry}}
\newcommand{\galmod}{\textsc{Galmod}}
\newcommand{\gipsy}{\textsc{GIPSY}}

\title[Inference for Gas Kinematics]{The SAMI Galaxy Survey: Bayesian Inference for Gas Disk Kinematics using a Hierarchical Gaussian Mixture Model}

\author[Varidel et al.]{Mathew R. Varidel$^{1,2,3}$\thanks{E-mail: mathew.varidel@sydney.edu.au},
Scott M. Croom$^{1,2,3}$,
Geraint F. Lewis$^{1}$,
Brendon J. Brewer$^{4}$,
\newauthor
Enrico M. Di Teodoro$^{5}$,
Joss Bland-Hawthorn$^{1, 3}$,
Julia J. Bryant$^{1, 2, 3, 6}$,
\newauthor
Christoph Federrath$^{5}$,
Caroline Foster$^{1, 3}$,
Karl Glazebrook$^{3, 7}$,
Michael Goodwin$^{8}$,
\newauthor
Brent Groves$^{2, 3, 5}$,
Andrew M. Hopkins$^{9}$,
Jon S. Lawrence$^{9}$,
\'Angel R. L\'opez-S\'anchez$^{9, 10}$,
\newauthor
Anne M. Medling$^{5, 11}$\thanks{Hubble Fellow},
Matt S. Owers$^{10, 12}$,
Samuel N. Richards$^{13}$
Richard Scalzo$^{14}$,
\newauthor
Nicholas Scott$^{1, 2, 3}$,
Sarah M. Sweet$^{3, 8}$,
Dan S. Taranu$^{2, 15, 16}$,
Jesse van de Sande$^{1, 3}$
\\
$^{1}$Sydney Institute for Astronomy (SIfA), School of Physics, A28, The University of Sydney, NSW 2006, Australia\\
$^{2}$ARC Centre of Excellence for All-Sky Astrophysics (CAASTRO)\\
$^{3}$ARC Centre of Excellence for All Sky Astrophysics in 3 Dimensions (ASTRO 3D)\\
$^{4}$Department of Statistics, The University of Auckland, Private Bag 92019, Auckland 1142, New Zealand \\
$^{5}$Research School of Astronomy and Astrophysics, Australian National University, Canberra, ACT 2611, Australia \\
$^{6}$Australian Astronomical Optics, AAO-USydney, School of Physics, University of Sydney, NSW 2006, Australia \\
$^{7}$Centre for Astrophysics and Supercomputing, Swinburne University of Technology, PO Box 218, Hawthorn, VIC 3122 \\
$^{8}$Australian Astronomical Observatory, 105 Delhi Rd, North Ryde, NSW 2113, Australia \\
$^{9}$Australian Astronomical Optics, Faculty of Science and Engineering, Macquarie University, 105 Delhi Rd, North Ryde, NSW 2113, Australia \\
$^{10}$Department of Physics and Astronomy, Macquarie University, NSW 2109, Australia \\
$^{11}$Ritter Astrophysical Research Center, University of Toledo, Toledo, OH 43606, USA \\
$^{12}$Astronomy, Astrophysics and Astrophotonics Research Centre, Macquarie University, Sydney, NSW 2109, Australia \\
$^{13}$SOFIA Science Center, USRA, NASA Ames Research Center, Building N232, M/S 232-12, P.O. Box 1, Moffett Field, CA 94035-0001, USA \\
$^{14}$Centre for Translational Data Science, University of Sydney, Darlington NSW 2008, Australia \\
$^{15}$International Centre for Radio Astronomy Research, University of Western Australia, 35 Stirling Highway, Crawley WA 6009, Australia \\
$^{16}$Department of Astrophysical Sciences, Princeton University, 4 Ivy Lane, Princeton, NJ 08544, USA \\
}

\date{Accepted XXX. Received YYY; in original form ZZZ}

\pubyear{2019}

\begin{document}
\label{firstpage}
\pagerange{\pageref{firstpage}--\pageref{lastpage}}
\maketitle

\begin{abstract}
We present a novel Bayesian method, referred to as \blobby{}, to infer gas kinematics that mitigates the effects of beam smearing for observations using Integral Field Spectroscopy (IFS). The method is robust for regularly rotating galaxies despite substructure in the gas distribution. Modelling the gas substructure within the disk is achieved by using a hierarchical Gaussian mixture model. To account for beam smearing effects, we construct a modelled cube that is then convolved per wavelength slice by the seeing, before calculating the likelihood function. We show that our method can model complex gas substructure including clumps and spiral arms. We also show that kinematic asymmetries can be observed after beam smearing for regularly rotating galaxies with asymmetries only introduced in the spatial distribution of the gas. We present findings for our method applied to a sample of 20 star-forming galaxies from the SAMI Galaxy Survey. We estimate the global H$\alpha$ gas velocity dispersion for our sample to be in the range $\bar{\sigma}_v \sim $[7, 30] km s$^{-1}$. The relative difference between our approach and estimates using the single Gaussian component fits per spaxel is $\Delta \bar{\sigma}_v / \bar{\sigma}_v = - 0.29 \pm 0.18$ for the H$\alpha$ flux-weighted mean velocity dispersion.

% This is a simple template for authors to write new MNRAS papers.
% The abstract should briefly describe the aims, methods, and main results of the paper.
% It should be a single paragraph not more than 250 words (200 words for Letters).
% No references should appear in the abstract.

\end{abstract}


%%%%%%%%%%%%%%%%%%%%%%%%%%%%%%%%%%%%%%%%%%%%%%%%%%

%%%%%%%%%%%%%%%%% BODY OF PAPER %%%%%%%%%%%%%%%%%%

\section{Introduction}
\label{sec:Introduction}

Accurately estimating the intrinsic gas kinematics is vital to answer specific science questions. For example, an open question remains about the drivers of turbulence within disk galaxies \citep[eg.][]{Tamburro2009,Federrath2017IAUS}. There is much evidence for higher velocity dispersions in $z>1$ galaxies compared to nearby galaxies \citep{Epinat2010,ForsterSchreiber2009,Genzel2006,Law2007,Wisnioski2011}. While the physical drivers of turbulence are not well understood, possibilities include one or more of the following; unstable disk formation \citep{Bournaud2010}, Jeans collapse \citep{Aumer2010}, star-formation feedback processes \citep{Green2010,Green2014}, cold-gas accretion \citep{Aumer2010}, ongoing minor mergers \citep{Bournaud2009}, interactions between clumps  \citep{Dekel2009a,Dekel2009b,Ceverino2010}, interactions between clumps and spiral arms \citep{Dobbs2007}, or interactions between clumps and the interstellar medium \citep{Oliva-Altamirano2018}.

To gain a better understanding of the drivers of gas turbulence within the disk, it is important to accurately determine the intrinsic velocity dispersion of the galaxy. However, a known issue of observations using spatially resolved spectroscopy is beam smearing. Beam smearing is the effect of spatially blurring the flux profile due to the atmospheric seeing. For observations using spectroscopy, beam smearing acts to spatially blend spectral features. The blending of spectral features at different Line of Sight (LoS) velocities acts to flatten the observed velocity gradient and increase the observed LoS velocity dispersion. For single-component disk models, this has been shown to greatly exacerbate the observed LoS velocity dispersion in the middle of the galaxy \citep{Davies2011}.

Several heuristic approaches have been used to estimate the intrinsic velocity dispersion of a galaxy. A popular approach is to estimate the velocity dispersion away from the centre of the galaxy \citep[eg.][]{Johnson2018}. Another approach is to apply corrections to the observed velocity dispersion as a function of properties that exacerbate the effect of beam smearing such as the seeing width and rotational velocity \citep{Johnson2018}. The local velocity gradient \citep{Varidel2016} has also been used to ignore spaxels with high local velocity gradient \citep{Zhou2017,Federrath2017} as well as provide corrections for the global \citep{Varidel2016} and local velocity dispersion \citep{Oliva-Altamirano2018}. 

Forward modelling approaches have also been used to simultaneously model the flux and kinematic profiles. In these algorithms, a 3D modeled cube is constructed for the galaxy and then spatially convolved per spectral slice to simulate the effect of beam smearing. The convolved cube is compared to the observed data. In this way, the galaxy properties are fitted to the original data while accounting for the effects of beam smearing. There are several publicly available cube-fitting algorithms designed for optical observations known to the authors. Those are \galpak{} \citep{Bouche2015}, \gbkfit{} \citep{Bekiaris2016}, and \bbarolo{} \citep{DiTeodoro2015}.

\galpak{} and \gbkfit{} assume parametric radial flux and velocity profiles with constant velocity dispersion. These algorithms have been used to infer the intrinsic global velocity dispersion and bulk rotation properties \citep[eg.][]{Contini2016,Oliva-Altamirano2018}. However, due to the parametric construction of the galaxy models, the residuals often exhibit significant substructure. This will usually be dominated by the gas distribution as it often exhibits more complex structure than the idealised radial profiles.

An implementation of non-parametric radial profiles has been constructed in tilted ring models. These models decompose the galaxy into a series of rings each with independent flux and kinematic properties. Tilted ring models are appropriate for analysing galaxies that are well represented by non-parametric radial profiles. In particular, they produce exquisitely detailed non-parametric radial profiles for high-resolution data \citep[eg. Fig. 4,][]{DiTeodoro2015}.

A pioneering 3D tilted ring model was implemented in \galmod{} \citep{Sicking1997} in the Gronigen Image Processing SYstem \citep[\gipsy,][]{vanderHulst1992}. Examples of modern implementations of tilted-ring models are \bbarolo{} and \tirific. \tirific{} has received considerable development allowing for increased flexibility on a standard tilted ring model. However, it has solely been used for HI radio observations. This is at least partially due it assuming the spectral dimension is frequency. While it would be possible to transform the optical wavelength dimension of the data to frequency for use in \tirific{}, we are not aware of researchers that have used \tirific{} on optical data. Instead, \bbarolo{} has been used on both optical \citep[eg.][]{DiTeodoro2016,DiTeodoro2018} and radio observations \citep[eg.][]{Iorio2017}.

A typical assumption used in previous methods is that the gas substructure can be well modelled using a radial profile. However, the distribution of gas within a galaxy is often more complex including rings, spiral arms, or individual clumps. In this paper, we will outline a 3D method to model the gas distribution and kinematic profiles robustly despite substructure of the gas distribution within the disk. This algorithm is inspired by the works of \citet{Brewer2011Lensing,Brewer2016Lensing}, who modelled the photometry of lensed galaxies with substructure by decomposing galaxies into a number of blobs using mixture models of a positive definite basis function. Our method (referred to as \blobby) decomposes the gas distribution into a mixture model of a positive definite basis function while simultaneously fitting the gas kinematics. Our method assumes radial velocity and velocity dispersion profiles across the galaxy.

The outline of this paper is as follows. In Section \ref{sec:ModelDescription} we will frame the inference problem in terms of Bayesian reasoning and describe the model parameterisation. In Section \ref{sec:TestModelling} we will discuss applications of our method to several toy data sets. In Section \ref{sec:SAMIModelling} we will apply the method to a sample of galaxies from the SAMI Galaxy Survey. In Section \ref{sec:Discussion} we will discuss the implications of our results. We then make our concluding statements in Section \ref{sec:Conclusions}.

\section{Model Description}
\label{sec:ModelDescription}

The problem of inferring the underlying galaxy properties can be formulated within the Bayesian framework as an inference for the galaxy parameters ($\mathbf{G}$), convolution parameters from the seeing and instrumental broadening ($\bm{\Sigma}$), and any systematic effects ($\mathbf{S}$) given some data ($D$),
\begin{align}
    p(\mathbf{G}, \Sigma, \mathbf{S} | D) 
        & \propto 
            p(\mathbf{G}, \Sigma, \mathbf{S}) 
            p(D | \mathbf{G}, \Sigma, \mathbf{S}) \\
        & \propto 
            p(\Sigma) 
            p(\mathbf{S} | \Sigma) 
            p(\mathbf{G} | \Sigma, \mathbf{S}) 
            p(D | \mathbf{G}, \Sigma, \mathbf{S}).
    \label{eq:genbinf}
\end{align}
Bayes' theorem relates the inference for the parameters $\mathbf{G}$, $\Sigma$, and $\mathbf{S}$ to our prior understanding in $p(\mathbf{G}, \bm{\Sigma}, \mathbf{S})$ and the data using the likelihood function, $p(D | \mathbf{G}, \bm{\Sigma}, \mathbf{S})$. All galaxy inferences can be summarised in this way. 

In this work, we will assume that the convolution parameters are known. That is, $p(\bm{\Sigma})$ is a delta function that peaks at the assumed convolution parameters. The Point Spread Function (PSF), representing the seeing, is typically estimated by modelling stars that are observed at the same time as the galaxies. Whereas the instrumental broadening is estimated by taking calibrations of the spectrograph using arc frames. Assuming that the convolution parameters are known will probably result in narrower posterior distributions than if we propagated our uncertainty in the convolution parameters. 

Furthermore, we  only consider systematic effects that are independent of the galaxy parameterisation. Making the above assumptions, we approximate the problem represented in equation (\ref{eq:genbinf}) to,
\begin{align}
    p(\mathbf{G}, \mathbf{S} | D, \Sigma) 
        &\propto 
            p(\mathbf{G}, \mathbf{S}) 
            p(D | \mathbf{G}, \bm{\Sigma}, \mathbf{S}) \\
        & \propto 
            p(\mathbf{G}) 
            p(\mathbf{S}) 
            p(D | \mathbf{G}, \bm{\Sigma}, \mathbf{S}).
    \label{eq:genbinf2}
\end{align}
The following sections will outline the assumptions made about the parameterisation of $\mathbf{G}$, $\bm{\Sigma}$, and $\mathbf{S}$.

\subsection{Galaxy parameterisation ($\mathbf{G}$)}
\label{sec:model} % used for referring to this section from elsewhere

Our choice of galaxy parameterisation is constructed with the aim to model the gas distribution and kinematics for a wide range of regularly rotating galaxies. We parameterise the gas distribution with respect to a single emission line. 

A simplistic prior assumption for the gas distribution of a galaxy, is that it consists of an unknown number of gas clouds that are gravitationally bound. The gas distribution will be centred and rotate around a single kinematic centre. The velocity profile is assumed to be radial with a gradient that is steep near the kinematic centre and plateaus at increasing radius. The velocity dispersion profile is assumed to follow a smoothly varying radial profile across the galaxy.

We will now describe the parameterisation of the above prior assumption in accordance with Bayes' theorem. Note that we also describe the joint prior distribution including the assumed constants, parameters, hyper-parameters, and data in Table \ref{tab:prior_table}.

\begin{table*}
	\centering
	\caption{
    The hyperparameters, parameters, and data (i.e. all of the quantities involved in the inference), along with the prior distributions for each quantity. Taken together, these specify the joint prior distribution for the hyperparameters, parameters, and data, from which we obtain the posterior distribution. Where parameters are assumed to be known we represent the prior as a Dirac delta function with a user-input defined as $\mathcal{U}$. The notation $T(a, b)$ (written after a probability distribution) denotes truncation to the interval $[a, b]$. \texttt{ImageWidth} and \texttt{PixelWidth} refer to the geometric mean of the spatial dimensions for the cube and a single pixel, respectively. Note that flux units are 10$^{-16}$ erg s$^{-1}$ cm$^{-2}$.
    }
	\begin{tabular}{lcr} % three columns, alignment for each
		\hline
			\textbf{Quantity} & \textbf{Meaning} & \textbf{Prior} \\		
		%\hline	
		%\hline
		%	\multicolumn{3}{l}{\textbf{Constants}} \\
		%\hline
            	%	$A_{k, \text{PSF}}$
              	%	  & Weight for each Gaussian representing the PSF 
              	% 	  &  Assumed to be known \\			
		%	$\text{FWHM}_{k, \text{PSF}}$ 
		%	  & Seeing FWHM for each Gaussian representing the PSF 
		%	  &  Assumed to be known \\
		%	$\text{FWHM}_{\text{lsf}}$
		%	  & Instrumental broadening 
		%	  & Assumed to be known  \\
		\hline
		    \multicolumn{3}{l}{\textbf{Galaxy coordinate system ($\mathbf{C}$)}} \\
		\hline
			$x_c$ 
			  & $x$-coordinate for centre of galaxy 
			  & Cauchy(\texttt{XImageCentre}, $0.1 \times \texttt{ImageWidth})
			    T(x_\text{min}, x_\text{max}$) \\
			$y_c$
			  & $y$-coordinate for centre of galaxy
			  & Cauchy(\texttt{YImageCentre}, $0.1\times\texttt{ImageWidth})     T(y_\text{min}, y_\text{max}$) \\
			$\theta$
			  & Galaxy semi-major axis angle (anti-clockwise w.r.t. East) 
			  & Uniform(0, 2$\pi$) \\
		         $i$ 
        			  & Galaxy inclination ($i$ = 0 for face-on) 
          		  &  $\delta (i - \mathcal{U})$ \\
		\hline
		    \multicolumn{3}{l}{\textbf{Number of blobs}} \\
		\hline
			$N$ 
			  & Number of blobs comprising the galaxy 
			  & Loguniform$\{1, 2, ..., 300 \}$ \\
		\hline
			\multicolumn{3}{l}{\textbf{Blob hyperparameters ($\bm{\alpha}$)}} \\
		\hline
			$\mu_r$ 
			  & Typical distance of blobs from ($x_c, y_c$) 
			  & Loguniform(0.03$''$, 30$''$) \\
			$\mu_{F}$ 
			  & Typical flux of blobs 
			  & Loguniform($10^{-3}, 10^{3}$) \\
			$\sigma_{F}$
			  & Deviation of log flux from $\mu_{F}$
			  & Loguniform(0.03, 3) \\
    		$W_{\text{max}}$ 
    		  & Maximum width of blobs 
    		  & Loguniform(\texttt{PixelWidth}, 30$''$) \\
			$q_{\text{min}}$
			  & Cutoff axis ratio
			  & Uniform(0.2, 1) \\
		\hline
			\multicolumn{3}{l}{\textbf{Blob parameters ($\mathbf{B}_j$)}} \\
		\hline
			$F_j$
			  & Integrated flux 
			  & Lognormal($\mu_F, \sigma^2_F$) \\
			$r_j$
			  & Distance of centre from ($x_c, y_c$) 
			  & Exponential($\mu_r$) \\
			$\theta_j$
			  & Polar angle of centre w.r.t. $\theta$ 
			  & Uniform(0.0, 2$\pi$) \\
            $w_j$ 
              & Width of blob 
              & Loguniform(\texttt{PixelWidth}, $W_{\text{max}}$) \\
			$q_j$ 
			  & Axis ratio ($q = b/a$) 
			  & Triangular($q_{\text{min}}$, 1) \\
			$\phi_j$
			  & Orientation angle (anti-clockwise w.r.t. $\theta + \theta_j$)
			  & Uniform(0, $\pi$) \\
		\hline
			\multicolumn{3}{l}{\textbf{Velocity profile parameters ($\mathbf{V}$)}} \\
		\hline
        	$v_\text{sys}$
        	  & Systemic velocity 
        	  & Cauchy(0 km s$^{-1}$, 30 km s$^{-1}$)
        	    $T$(-150 km s$^{-1}$, 150 km s$^{-1}$) \\
			$v_c$
			  & Asymptotic velocity 
			  & Loguniform(40 km s$^{-1}$, 400 km s$^{-1}$) \\
			$r_t$
			  & Turnover radius for velocity profile
			  & Loguniform(0.03$''$, 30$''$) \\
			$\gamma_v$
			  & Shape parameter for velocity profile
			  & Loguniform(1, 100) \\ 
            $\beta_v$
              & Shape parameter for velocity profile 
              & Uniform(-0.75, 0.75) \\
		\hline
			\multicolumn{3}{l}{\textbf{Velocity dispersion profile parameters ($\bm{\Sigma}_\mathbf{V}$)}} \\
		\hline
            $\sigma_{v, 0}$
              & Velocity dispersion at the kinematic centre 
              & Loguniform(1 km s$^{-1}$, 200 km s$^{-1}$) \\
            $\sigma_{v, 1}$
              & Log velocity dispersion gradient
              & Normal(0, 0.2$^2$) \\
		\hline
			\multicolumn{3}{l}{\textbf{Convolution parameters ($\bm{\Sigma}$)} }\\
		\hline
    		$A_{k, \text{PSF}}$
      		  & Weight for each Gaussian representing the PSF 
      	 	  &  $\delta (A_{k, \text{PSF}} - \mathcal{U})$ \\	
			$\text{FWHM}_{k, \text{PSF}}$
			  & Seeing FWHM for each Gaussian representing the PSF 
			  &  $\delta (\text{FWHM}_{k, \text{PSF}} - \mathcal{U})$ \\
			$\text{FWHM}_{\text{lsf}}$
			  & Instrumental broadening 
			  &  $\delta (\text{FWHM}_{\text{lsf}} - \mathcal{U})$ \\
		\hline
			\multicolumn{3}{l}{\textbf{Systematic parameters ($\mathbf{S}$)}} \\
		\hline
			$\sigma_0$
			  & Constant Gaussian noise component
			  & Loguniform($10^{-12}$, 10) \\
		\hline
			\multicolumn{3}{l}{\textbf{Data ($D$)}} \\    
		\hline
        	$D_{ijk}$
        	  & Flux for each velocity bin 
        	  & Normal($M_{ijk}$, $\sigma^2_{\text{obs}}$ + $\sigma^2_0$)  
	\end{tabular}
	\label{tab:prior_table}
\end{table*}

\subsubsection{The galaxy coordinate system}
\label{sec:GlobalModel}

The galaxy coordinate system is described by a kinematic centre at $(x_c, y_c)$, an inclination angle $i$, and the semi-major axis position angle $\theta$. This describes a thin plane for the gas to lie in. The set of parameters required to define the coordinate system are referred to as $\mathbf{C}$. The prior distribution for each parameter is assumed to be independent such that,
\begin{equation}
	p(C) = p(x_c) p(y_c) p(i) p(\theta)
	\label{eq:coordprior}
\end{equation}

The kinematic centre of the galaxy is typically in the centre of the Field-of-View (FoV). We weakly incorporate this information by placing a wide-tailed Cauchy distribution centred in the middle of the image with a Full-Width Half-Maximum (FWHM) of $0.1\times \texttt{ImageWidth}$. \texttt{ImageWidth} is defined to be the geometric mean length of the FoV. The prior distribution for the kinematic centre is truncated such that it cannot lie outside of the FoV.

We assume that the kinematic position angle follows a uniform distribution in the range $\theta \in [0, 2\pi]$. The inclination angle is typically constrained by the observed morphology and the kinematic profiles. However, it is often not possible to observe the full extent of the galaxy in IFS surveys. For example, a typical galaxy observed in the SAMI Galaxy Survey, which we will be using to test our methodology, is observed out to $\sim 2 R_e$, where $R_e$ is the half-light radius. This limits our ability to infer the inclination from the observed gas distribution. The LoS kinematic profiles are known to be approximately degenerate for varying inclination angles as well \citep[eg. Fig. 9,][]{Glazebrook2013}. We did test our methodology with a uniform prior for the inclination angle in the range $i \in [0, \pi/2]$. However, when applying our methodology to the sample galaxies in Section \ref{sec:SAMIModelling}, we found that the inferred inclination angle could differ significantly from the estimated inclination angle when converting the observed ellipticity to an inclination angle assuming a thin disk. With this in mind, we assume that the inclination can be estimated from previous observations of the same galaxy with a wider FoV. The inclination is then set as a constant. The inclination and kinematic position angle are incorporated into the LoS velocity profile and define a plane that the gas lies in.

Setting the inclination angle as a constant will have several implications for our inferences. The inferred posterior distributions will probably be narrower than if we incorporated our uncertainty of the inclination angle into our model parameterisation. Also, the effect of beam smearing on kinematic properties is a function of the LoS velocity profile which is affected by the inclination angle assumption. As such, we will introduce a systematic bias when our assumptions about the inclination are incorrect.

\subsubsection{The spatial gas distribution}
\label{sec:GasDistributionModel}

To incorporate our prior understanding within the galaxy parameterisation, we decomposed the gas distribution into a sum of positive definite basis functions. We use positive definite basis functions as the integrated flux of a gas cloud should always be positive. Decomposing the gas distribution into a sum of positive definite basis functions is an approach to model complex structures such as spirals, rings, and clumps that are observed in galaxies. We refer to each component as a `blob'.

We do not claim that a single blob represents an individual gas cloud. This is due to the following:
\begin{itemize}
    \item The resolution of the data in many IFS studies is typically too low to resolve individual gas clouds.
    \item The choice of parameterisation for the positive definite basis function will lead to more or less blobs. This is due to the shape of the blob not perfectly matching the individual gas cloud. As such, several blobs may be required to model the shape of the gas cloud.
\end{itemize}
There are cases where an individual blob or a set of blobs may be assigned a particular classification such as an individual clump, spiral arm, or ring. However, such processing of the model output must be performed by the user after the modelling has been completed. For the majority of cases, the individual blobs should be seen as nuisance parameters. The primary reason for using blobs is to construct a flexible model of the gas distribution, rather than to derive properties of individual gas clouds.

There have been previous 3D approaches that decomposed galaxies into a series of sources (ie. clouds or blobs). An example of this are the Monte Carlo integration techniques used in tilted ring models such as \galmod{} \citep{Sicking1997}. In these algorithms, the 3D tilted ring model is integrated using Monte Carlo sampling of point sources within a ring with a given gas column density and kinematics. However, the primarily goal is not to derive the individual parameters of the clouds, but rather to perform the integration of the 3D tilted ring model.

An alternative flexible approach, that has been applied to lensing data, is to use pixelated flux profiles. In these models, each pixel has an independent flux value. The pixelated flux profile is often regularised such that the resulting profile is smooth \citep[eg.][]{Suyu2006}. The advantage of this approach is that it can theoretically model any flux distribution at the observed scale, prior to performing the convolution. The disadvantage of the pixelated approach, is that the prior distribution assigns high prior probability to flux profiles that look like noise and the regularisation approach typically does not enforce the flux to be positive definite \citep{Brewer2011Lensing}. As such, we have chosen to use the approach of modelling the gas distribution using a sum of positive definite basis functions.

We chose a Gaussian basis function where the integrated flux for each blob is always positive. Using a Gaussian basis function to represent the spatial gas distribution is not the only possibility. For example, generic S\`{e}rsic profiles and quadratic polynomials with negative curvature calculated where the flux is positive have been used to model lensed galaxies by \citet{Brewer2011Lensing,Brewer2016Lensing}. Other paramaterisations of positive definite functions would also be feasible.

Each blob is defined by a set of parameters $\mathbf{B}_j$ that describe its integrated flux ($F_j$), central position ($r_j, \theta_j$) with respect to the galaxy centre $(x_c, y_c)$ and semi-major axis position angle ($\theta$), width ($w_j$), axis ratio ($q_j = b/a$), and orientation ($\phi_j$) with respect to $\theta + \theta_j$. The spatial component of the blob flux is then,
\begin{equation}
    F(x', y') = \frac{F_j}{2 \pi w_j^2} \exp \bigg( 
            -\frac{1}{2w_j^2} \bigg(  q_j x'^2+  \frac{y'^2}{q_j} \bigg) \bigg).
    \label{eq:sflux}
\end{equation}
The coordinate system $(x', y')$ is transformed with respect to the galaxy coordinate system defined by $\mathbf{C} = \{ x_c, y_c, i, \theta \}$ and subsequently rotated with respect to the blob orientation ($\phi_j$). To construct the flux map in the original coordinate system (ie.\ $F(x, y)$), we calculate the flux per spaxel in the rotated coordinates and sum the flux contribution for each blob.

The blob parameters $F_j$, $r_j$, $w_j$, and $q_j$ are hierarchically constrained. Hierarchical Gaussian mixture models refer to models that are a sum of Gaussians where the Gaussian parameters are hierarchically constrained. For a hierarchical Gaussian mixture model, a joint prior is constructed for the Gaussian parameters $\{ \mathbf{B}_j \}^N_{j=1}$ for $N$ Gaussians conditional on a set of hyperparameters $\bm{\alpha}$ (ie.\ the parameters for the prior distribution). The joint prior distribution for $N$ Gaussians is then described as,
\begin{equation}
    p(\bm{\alpha}, \{\mathbf{B}_j\}^{N}_{j=1}) = 
        p(\bm{\alpha}) \prod^{N}_{j=1} p(\mathbf{B}_j | \bm{\alpha}).
    \label{eq:hbm}
\end{equation}
Where $p(\bm{\alpha})$ refers to the prior distribution for the hyperparameters. The prior distribution for the blob parameters $\mathbf{B}_j$ are dependent on the hyperparameters encoded in $p(\mathbf{B}_j | \bm{\alpha})$.

The number of Gaussians required to adequately model the gas distribution is unknown prior to performing the inference. We can explicitly incorporate this within the joint prior distribution such that,
\begin{align}
      p(N, \bm{\alpha}, \{\mathbf{B}_j\}^N_{j=1}) 
        &= p(N) p(\bm{\alpha} | N) \prod^{N}_{j=1} 
            p(\mathbf{B}_j | \bm{\alpha}, N) \\
        &= p(N) p(\bm{\alpha}) \prod^{N}_{j=1} p(\mathbf{B}_j | \alpha).
    \label{eq:thbm}
\end{align}
The last step assumes the hyperparameters ($\bm{\alpha}$) and blob parameters $\{\mathbf{B}_j\}^{N}_{j=1}$ are independent from the number of Gaussians ($N$). We defined the prior distribution for the number of blobs $p(N)$ to be a loguniform distribution in the range \{1, 2, 3, ..., $N_\text{max}$ \}. We have set $N_\text{max} = 300$ for all examples in this paper. Given 6 parameters per blob and a potential for up to 300 blobs, the total number of parameters to describe the full set of Gaussians is between 6 -- 1,800.

Hierarchical Gaussian mixture models are preferred when the parameters for the Gaussians follow a prior distribution where the hyperparameters are unknown. In our case, the hyperparameters are descriptors for the distribution of blobs which are specific for the observed galaxy. In this way the galaxy shape, typical blob shape, and individual blob parameters are inferred simultaneously.

We assume the integrated flux of the blobs follows a lognormal distribution suggesting that the blob has a typical integrated flux ($\mu_F$) and deviation ($\sigma_F$). The lognormal distribution also ensures the integrated flux is positive.

The distance of the blobs ($r_j$) is assumed to follow an exponential distribution from the kinematic centre ($x_c$, $y_c$). This imparts a typical distance $\mu_r$ from the kinematic centre which is fitted per galaxy.

The width of the blobs ($w_j$) is assumed to follow a loguniform distribution. The choice of a loguniform distribution is chosen to avoid imparting a typical scale length as both disk and clumpy features may be required to model a given galaxy. The minimum width is defined by the \verb'PixelWidth' which is the geometric mean of the $x$ and $y$ dimensions for a pixel. Restricting the minimum width of the blobs has been incorporated for several reasons. It limits the problem of accurately integrating and spatially convolving blobs that are much smaller than the pixel width. It also limits the possibility of overfitting the gas substructure. The maximum width ($W_\text{max}$) is a free hyperparameter that is fitted for the galaxy.

The typical axis ratio ($q_j = b/a$) for a blob is also unknown prior to performing the inference. We chose a right-angled triangular prior distribution for $q_j$ of the form,
\begin{equation}
    p(q_j) =  \frac{ 2 (q_j - q_\text{min}) }{(1 - q_\text{min})^2}.
    \label{eq:qprior}
\end{equation}
The hyperparameter $q_\text{min}$ is the minimum axis ratio. This prior imparts a preference for circular Gaussians.

\subsubsection{The Velocity Profile}
\label{sec:VelocityModel}

In the spectral dimension, we assume a single Gaussian emission line component per spaxel. The mean position per spaxel describes the rotational velocity profile across the galaxy. We assumed a continuous velocity profile across the blobs with a mean LoS velocity defined by the \citet{Courteau1997} empirical model,
\begin{equation}
    v(r) = 
        v_\text{c}
        \frac{(1 + r_t/r)^\beta}{(1 + (r_t/r)^\gamma)^{1/\gamma}}
        \sin(i) \cos(\theta)
        + v_{\text{sys}}.
    \label{eq:vprof}
\end{equation}
$r$ is defined as the distance in the galaxy plane to the kinematic centre. $v_\text{sys}$ is a systemic velocity term, $v_c$ is the asymptotic velocity, and $r_t$ is the turnover radius. $\beta$ is a shape parameter that describes the gradient for $r > r_t$, where positive results in a decreasing velocity profile and negative results in a increasing profile. $\gamma$ describes how sharply the velocity profile turns over. We refer to the set of parameters describing the velocity profile as $\mathbf{V}$. 

The prior distribution for these parameters are assumed to be independent such that,
\begin{equation}
	p(\mathbf{V}) = p(v_\text{sys}) p(v_c) p(r_t) p(\beta) p(\gamma).
	\label{eq:vprior}
\end{equation}
It is assumed that the data cube is de-redshifted, but we allow for offsets for a non-zero systemic velocity by applying a prior that follows a wide-tailed Cauchy distribution with FWHM of 30 km s$^{-1}$ and is truncated to the interval [-150 km s$^{-1}$, 150 km s$^{-1}$]. For all examples explored in this paper, the systemic velocity was well within these ranges. However, the range can be increased to account for a greater offsets if required. 

\begin{figure}
    \begin{center}
        \includegraphics[
            width=0.5\textwidth,
            keepaspectratio=true,
            trim=0mm 2mm 0mm 0mm,
            clip=true]{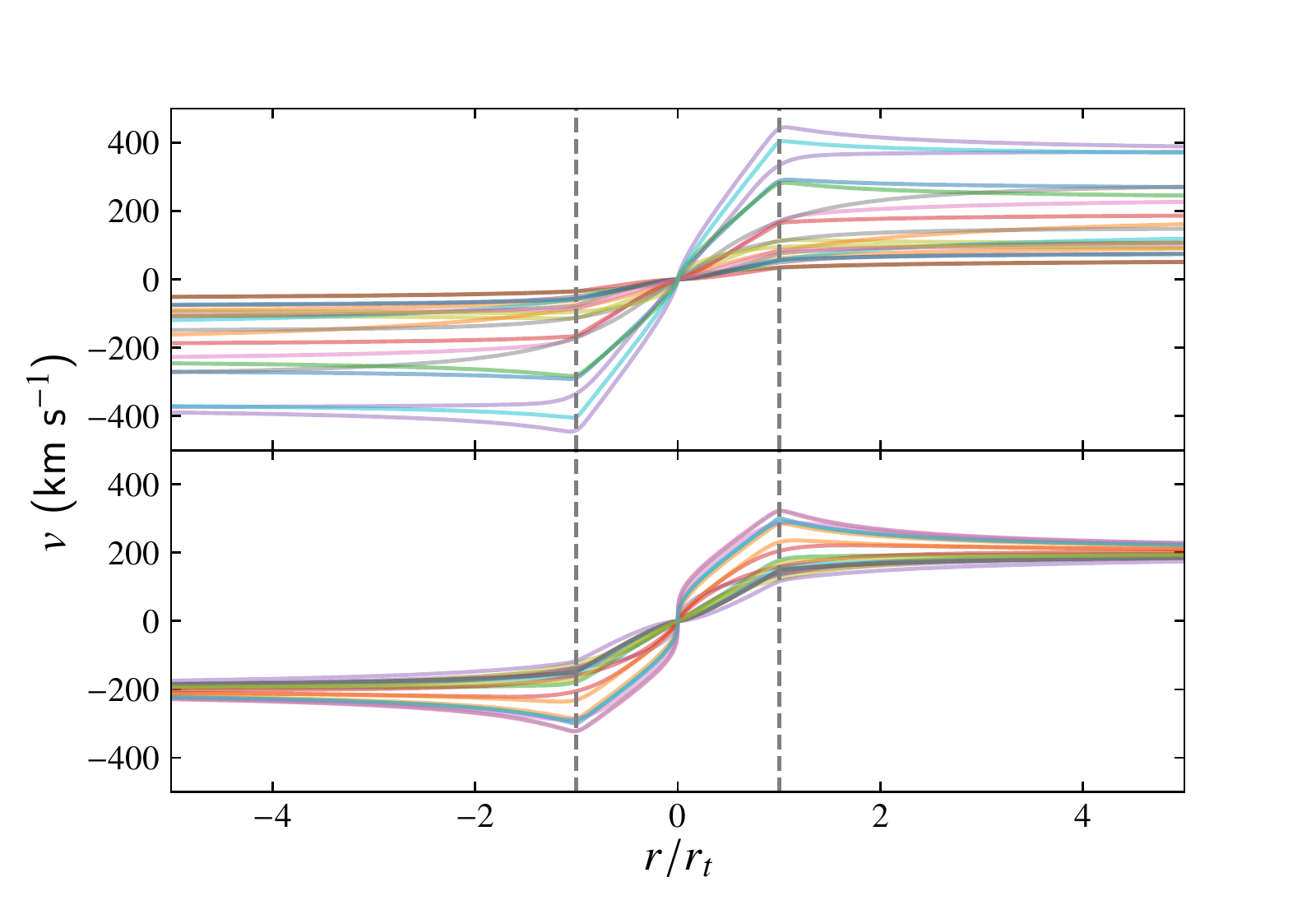}
    \end{center}
    \caption{
    Prior samples of the radial velocity profile. Samples where all velocity parameters vary except $v_\text{sys} = 0$ km s$^{-1}$ (top) and with $v_c = 200$ km s$^{-1}$ (bottom). Vertical lines indicate the turnover radius at $r = \pm r_t$. Our choice of priors for the velocity profile parameters were chosen to yield realistic radial velocity profiles.
    }
    \label{fig:vprior}
\end{figure}

The remaining parameters $v_c$, $r_t$, $\beta$, and $\gamma$ are set with limits that yield a reasonable prior distribution by observing samples of the profiles.  See Fig. \ref{fig:vprior} for velocity profiles using random samples from the prior for the velocity parameters. We assume loguniform prior for $v_c$ in the range [40 km s$^{-1}$, 400 km s$^{-1}$]. The lower bound of 40 km s$^{-1}$ for $v_c$ was adequate for the test galaxies in this paper, but it can be easily lowered to take into account a larger sample of galaxies. The turnover radius ($r_t$) is assumed to follow a loguniform distribution in the range [0.03$''$, 30$''$]. 

Our velocity profile assumption yields a reasonably flexible radial profile, but we do not claim that this represents all galaxy velocity profiles. In particular, warps and asymmetries are not taken into account. Further flexibility may be required when the method is applied to larger data sets.

\subsubsection{The velocity dispersion profile}
\label{sec:VelocityDispersionModel}

The width of the Gaussian in the spectral dimension describes velocity dispersion per spaxel. The velocity dispersion profile is assumed to be a log-linear radial profile of the form,
\begin{equation}
    \sigma_v(r) = \exp \bigg( 
        \log(\sigma_{v, 0}) + \sigma_{v, 1} r 
        \bigg).
    \label{eq:vdisp}
\end{equation}
Where $\sigma_{v, 0}$ represents the velocity dispersion at the kinematic centre ($x_c, y_c$) and $\sigma_{v, 1}$ represents the log radial velocity dispersion gradient. We refer to the set of parameters that describe the galaxy velocity dispersion profile as $\bm{\Sigma}_\mathbf{V}$. We used a log-linear profile such that $\sigma_v > 0$ at all radii. A disadvantage of this parameterisation is that for large $\sigma_{v, 1}$, the observed $\sigma_v$ can be much higher than is realistic. We use a normal prior distribution with mean 0 and variance $0.2^2$ for $\sigma_{v, 1}$ to limit unrealistically high velocity dispersion gradients. We assume independence of the prior distributions for $\bm{\Sigma}_\mathbf{V}$ such that,
\begin{equation}
	p(\bm{\Sigma}_\mathbf{V}) = p(\sigma_{v, 0}) p(\sigma_{v, 1}).
\end{equation}

During testing we also explored the possibility of having a single velocity dispersion per blob. While this would be ideal, it can lead to over-fitting systematics that have not been corrected for appropriately. In particular, blobs with unrealistically high velocity dispersion would often be required to account for systematic offsets in the continuum. This can occur in the log-linear model as well, but it is less affected due to the parameterisation across the galaxy. Therefore, we have opted for a simplified parametric model which is more robust but less flexible.

\subsubsection{The full galaxy parametersisation}
\label{sec:3dmodel}

The flux distribution including a Gaussian instrumental broadening ($\sigma_\text{lsf}$) within velocity space for a blob is defined as, 
\begin{equation}
    F(x, y, v) = \frac{F(x, y)}{\sqrt{2 \pi (\sigma^2_{v(r(x, y))} + \sigma^2_\text{lsf})}} 
    \exp{ \Bigg( 
        \frac
            {(v - v(r(x, y)))^2}
            { \sqrt{\sigma^2_{v(r(x, y))} + \sigma^2_\text{lsf}}}
        \Bigg)
        }.
    \label{eq:vflux}
\end{equation}
Equations \ref{eq:sflux}, \ref{eq:vprof}, \ref{eq:vdisp}, and \ref{eq:vflux} fully define the flux distribution of a blob for a given emission line for the spatial and velocity dimensions. The above model is converted from velocity to wavelength space such that the model can be compared to the data.

The full joint prior distribution for our galaxy model parameterisation is described as,
\begin{align}
    p(\mathbf{G}) 
     &= p(\mathbf{C}, \mathbf{V}, \bm{\Sigma}_\mathbf{V}, N, \bm{\alpha}, \{\mathbf{B}_j\}^{N}_{j=1}) \\
     &= p(\mathbf{C}) p(\mathbf{V}) p(\bm{\Sigma}_\mathbf{V}) p(N, \bm{\alpha}, \{\mathbf{B}_j\}^{N}_{j=1}) \\
     &= p(\mathbf{C})  p(\mathbf{V}) p(\bm{\Sigma}_\mathbf{V}) p(N) p(\bm{\alpha}) \prod^{N}_{j=1} p(\mathbf{B}_j | \alpha).
     \label{eq:gprior}
\end{align}
The first step expands the galaxy parameterisation ($\mathbf{G}$) to the sets of parameters describing the galaxy coordinate system ($\mathbf{C}$), velocity profile ($\mathbf{V}$), velocity dispersion profile ($\bm{\Sigma}_\mathbf{V}$), number of blobs ($N$), the hyperparameters for the blobs ($\bm{\alpha}$), and the blob parameters ($\{\mathbf{B}_j\}^{N}_{j=1}$). The second step assumes independence between the various parameter sets where applicable. The third step expands the joint prior for $N$, $\bm{\alpha}$, and $(\{\mathbf{B}_j\}^{N}_{j=1})$ to state the dependence of the blob parameters ($\{\mathbf{B}_j\}^N_{j=1}$) on the blob hyperparameters ($\bm{\alpha}$) as in Equation \ref{eq:thbm}.

\subsection{Sampling the prior for $\mathbf{G}$}
\label{sec:priorsamples}

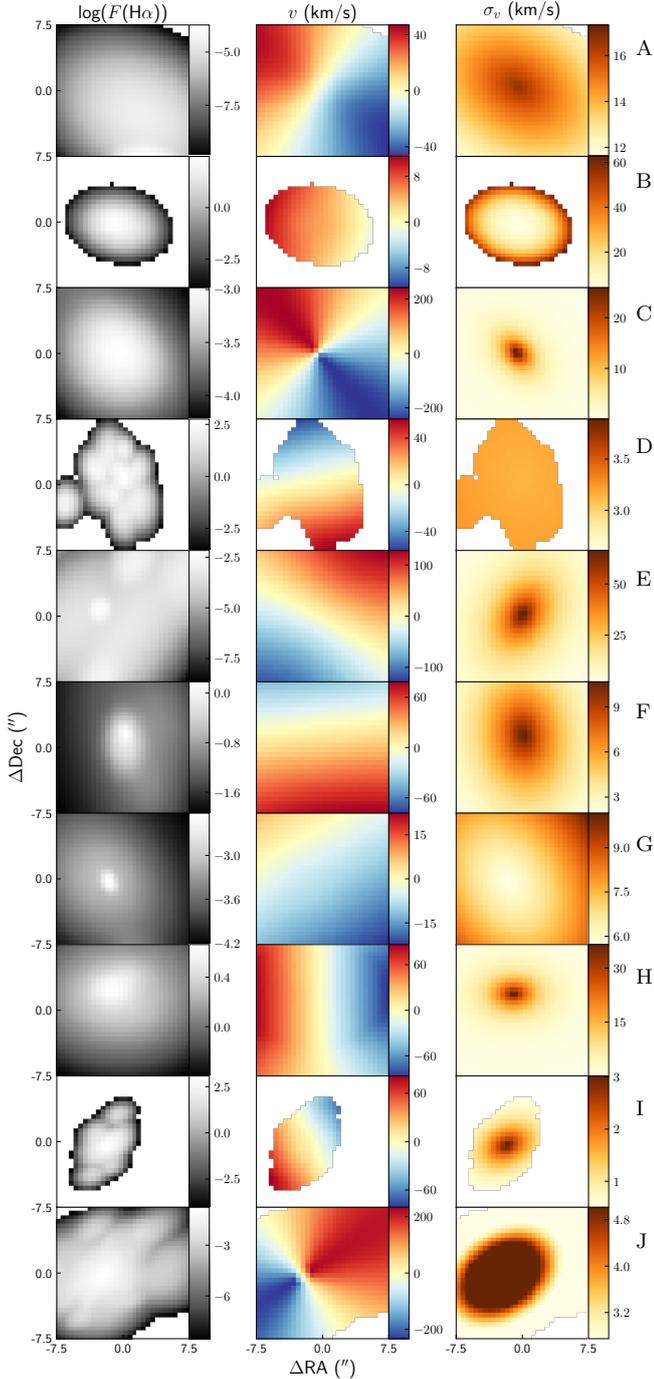
\begin{figure}
    \begin{center}
        \includegraphics[
            width=0.5\textwidth,
            keepaspectratio=true,
            trim=0mm 38mm 0mm 50mm,
            clip=true]{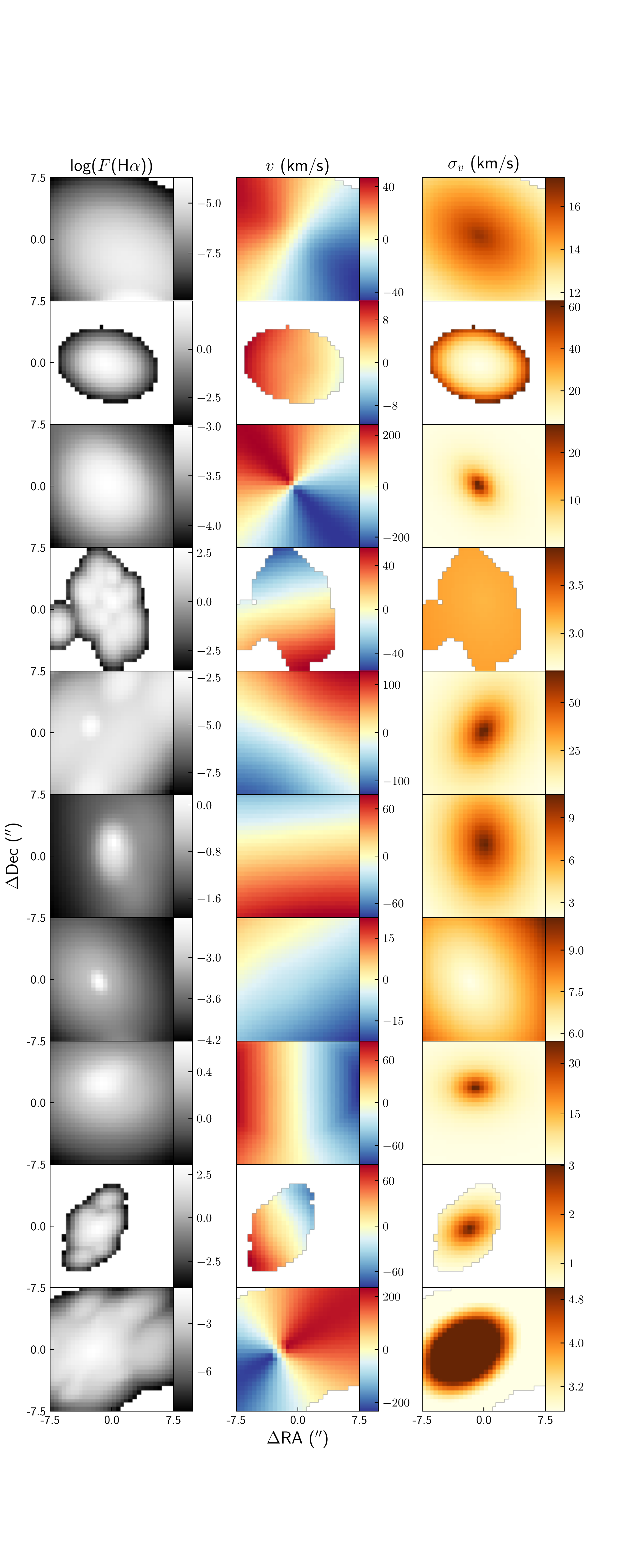}
        \put(-15, 500){A}
        \put(-15, 450){B}
        \put(-15, 400){C}
        \put(-15, 350){D}
        \put(-15, 300){E}
        \put(-15, 250){F}
        \put(-15, 200){G}
        \put(-15, 150){H}
        \put(-15, 100){I}
        \put(-15, 50){J}
    \end{center}
    \caption{
        2D maps of randomly drawn samples from the prior distribution for the H$\alpha$ flux (left), LoS velocity (middle), and LoS velocity dispersion (right). For illustrative purposes, we show samples with inclination $i = \pi/4$, systemic velocity $v_\text{sys} \in [-10 \text{ km s}^{-1}, 10 \text{ km s}^{-1}]$, and the kinematic centre $x_c, y_c \in [-3'', 3'']$. These maps show the flexibility of modelling the spatial gas distribution using a Gaussian mixture model. We also chose priors to yield realistic gas distributions and kinematic profiles.
        }
    \label{fig:priormaps}
\end{figure}

The galaxy model parameterisation is complex, including hierarchical constraints and a variable number of parameters dependent on the number of blobs. For such high dimensional model parameterisations, it is often difficult to gain an intuitive understanding of the prior distribution. A common approach to check that a complex prior distribution is reasonable, is to visually check randomly drawn samples from the prior. As an example of this approach, we show 2D maps for 10 randomly drawn samples from the joint prior distribution in Fig. \ref{fig:priormaps}.

The 2D maps are constructed with a $15''$ and $0.5''$ square FoV and pixel width. These limits were constructed with the SAMI Galaxy Survey in mind, which has a FoV with typical diameter of $\sim 15''$ and $0.5''$ square pixels. We set the inclination $i=\pi/4$. For illustrative purposes, we also limit the prior samples shown in Fig. \ref{fig:priormaps} such that $v_\text{sys} \in [-10 \text{ km s}^{-1}, 10 \text{ km s}^{-1}]$ and $x_c, y_c \in [-3'', 3'']$.

In all samples there is a clear photometric and kinematic centre.  These properties are constrained by the global parameters controlling the plane for the gas to lie in ($i$, $\theta$) as well as the centre and typical distance for the blob centres ($x_c$, $y_c$, $\mu_r$). Similarly, we avoid unusually shaped blobs by hierarchically constraining the width and axis-ratio of the blobs.

Several samples add increased complexity with centralised peaks (eg.\ F and G) and others with non-centralised clumps (eg.\ D, E, F, I, J). The most unusual clump is probably in D on the west-side of the image, but individual gas clumps similar to this are possible in real data \citep[eg.][]{Richards2014}.

The LoS velocity and velocity dispersion profiles are reasonable radial velocity profiles. Increased flexibility such as warps and asymmetries could be added to increase the realism of the profiles in the future.

We note that the prior distribution is a balance between flexibility and realism. As such, not all samples from the prior will represent realistic galaxies. Instead, the data is required to constrain the prior distribution via posterior sampling.

\subsection{PSF convolution}
\label{sec:convolution} % used for referring to this section from elsewhere

The PSF convolution kernel is assumed to be well represented by a decomposition of concentric circular 2D Gaussians. Each Gaussian is described by $\Sigma_k = \left\{A_{k, \text{PSF}}, \text{FWHM}_{k, \text{PSF}} \right\}$ corresponding to the weight and FWHM for the $k$-th component. Each Gaussian has the separability property such that it can be deconstructed into two orthogonal vectors. Therefore, the 2D convolution is performed by convolving consecutively along each axis. Linear convolution using this method scales as $\mathcal{O}( N_\text{col,image} N_\text{col,kern} + N_\text{row,image} N_\text{row,kern}$) for each Gaussian. Further speed-up is gained by only constructing each Gaussian kernel out to $2.12\times\text{FWHM}_\text{PSF}$, which is equivalent to 5$\sigma_\text{PSF}$.

Convolution is also a distributive operation. As such, we perform the convolution by each Gaussian component on the original image and then sum the convolved images. This method will scale linearly with the number of Gaussians required to model the kernel. We have only used 1--2 Gaussian components to represent the PSF as that was an acceptable number in our case.

In all examples in this paper, we have used representations of the kernel to be a Gaussian or Moffat profile. We do this as the pipeline for the SAMI Galaxy Survey provides estimates for the PSF for both the Gaussian and Moffat profiles. The PSF profile paramters are estimated by fitting observations of stars that have been taken simultaneously to observing the galaxies. In cases where the PSF is represented by a Moffat profile, we fit the 2D Moffat kernel with a sum of 2 Gaussians. The fitted parameters are then passed to the code implementation of our method.

\subsection{Data}
\label{sec:eval} % used for referring to this section from elsewhere

Our method assumes that the data cube has been isolated to a single emission line and the continuum has been subtracted. For optical IFS observations, this requires accurate modelling of the stellar continuum. In low signal-to-noise observations this may not be possible and thus signal-to-noise cuts of the data cube are required. While it may be ideal to parameterise the systematics in the continuum corrections, we avoided modelling the systematics to avoid introducing a high number of nuisance parameters to our model.

To isolate an emission line, typical optical IFS observations will need to be cut in the spectral dimension around the emission line of interest. This may be difficult in the spectral regions where there are multiple emission lines. In our examples, we will be focusing on the H$\alpha$ emission line at 6562.8 \AA{} which is adjacent to the two [NII] lines at  6548.1 \AA{} and 6583.1 \AA. Isolating the H$\alpha$ emission line from the surrounding [NII] lines may be impossible for galaxies with high LoS velocity dispersions. In such cases, it will be a requirement to model the [NII] lines as this will cause systematics which we have not taken into account in our current parameterisation. Adding the [NII] lines could be introduced to our method by modelling the [NII]/H$\alpha$ per blob, then constraining the doublet using the theoretical ratio between the lines.

To construct the likelihood function, we assume the data follows a normal distribution. The mean is equal to an input data cube file ($D_{ijk, \text{obs}}$). The variance is given by the sum of an input variance cube ($\sigma^2_{ijk, \text{obs}}$) and an additional constant variance ($\sigma^2_0$),
\begin{equation}
    \sigma^2_{ijk} = \sigma^2_{ijk, \text{obs}} + \sigma^2_0, 
    \label{eq:GenerativeSigma}
\end{equation}
$\sigma^2_0$ is a systematic noise parameter corresponding to $\mathbf{S}$ in our generic inference problem in Equation \ref{eq:genbinf2}. $\sigma^2_0$ helps take into account under-estimated variance within the continuum subtracted data cube and some systematics that may arise due to limitations in the galaxy model parameterisation. The additional variance term will not account for significant unresolved structures between the data and model. Under those circumstances, the posterior distributions can be systematically biased. 

The non-diagonal elements of the covariance cube have not been incorporated. Including the non-diagonal elements of the covariance would require an inversion of the covariance matrix which scales as $\mathcal{O}(n^3)$. Data cubes cut around H$\alpha$ typically have $\mathcal{O}(10^3)$ data points, which results in a highly time consuming calculation. As such, we have avoided implementing the covariance matrix in the likelihood function. The likelihood function is then given by,
\begin{equation}
    p(D | \mathbf{G}, \Sigma, \mathbf{S}) = 
    %\mathcal{L} = 
        \prod^{n_i} \prod^{n_j} \prod^{n_k} 
        \frac{1}{\sqrt{2 \pi \sigma^2_{ijk}}}
        \exp{ \bigg( 
            - \frac{(M_{ijk} - D_{ijk})^2}{2\sigma^2_{ijk}}
            \bigg)}.
    \label{eq:likelihood}
\end{equation}
where $M_{ijk}$ represents the model convolved by the PSF.

\subsection{Posterior sampling}
\label{sec:sampling} % used for referring to this section from elsewhere

The posterior density function (PDF) is defined by Equation \ref{eq:genbinf2}, where the joint prior for the galaxy parameterisation is given in Equation \ref{eq:gprior}, the prior for our systematic parameters is defined as $p(\mathbf{S}) = p(\sigma_0)$, and the likelihood function is given in Equation \ref{eq:likelihood}. Table \ref{tab:prior_table} also summarises the joint prior distribution and data. The galaxy model is described by 4 global parameters, 5 blob hyperparameters, 5 velocity parameters, 2 velocity dispersion parameters, 1 systematic noise parameter, and 6 blob parameters for $N$ blobs. For typical galaxies 10s--100s of blobs are required to sufficiently model the galaxy assuming our joint prior distribution. As such, the number of parameters required to model the galaxy is typically $\mathcal{O}(100)$, making this a high parameter model. It is also required to fit both the number of blobs as well as the parameters for those blobs.

With these requirements in mind, we use \dnest{} \citep{Brewer2011DNEST,Brewer2018DNEST4}. \dnest{} expands the nested sampling aglorithm \citep{Skilling2004} by constructing future levels via a multi-level exploration of the posterior density function. The multi-level exploration is performed using an implementation of the Metropolis algorithm in the the Markov-Chain Monte-Carlo (MCMC) class. \dnest{} is typically more robust to local maxima as it has the ability to walk up and down nested sampling levels to explore the posterior distribution. Furthermore, as \dnest{} is a nested sampling algorithm it can be used to calculate the evidence $Z$ (ie. the normalisation constant for a given model), and subsequently perform model comparison.

\dnest{} also has an in-built reversible jump object \citep{Brewer2014}. A reversible jump is a proposal step that allows for a change in components. We use this to propose steps that add or remove blobs such that we can perform posterior sampling for the number of blobs ($N$). An inference problem with a varying number of components is referred to as transdimensional inference. Such problems are notoriously difficult to explore, but \dnest{} has been used to successfully perform inferences on such problems as modelling lensed galaxies with a variable number of blobs \citep{Brewer2011Lensing,Brewer2016Lensing}, similar to our approach. Other applications within astronomy have been to estimate the number of stars in a crowded stellar field \citep{Brewer2013} and modelling star-formation histories \citep{Walmswell2013}.

\section{Testing the Method}
\label{sec:TestModelling}

The remaining sections of this paper are devoted to demonstrating the methodology on a number of examples. We have tested the method on idealised toy models and real data. In this section, we will describe the applications of our method applied to a set of toy models.

\subsection{Simple toy models}
\label{sec:tm} % used for referring to this section from elsewhere

The toy models were constructed as a thin disk with an exponential flux profile. The velocity dispersion was set to a constant across the disk. We used an Universal Rotation Curve \citep[URC,][]{Persic1996} to model the velocity profile.

The URC was chosen as this profile relates the flux profile to the velocity profile via the parameter $v(R_\text{opt})$, where $R_\text{opt}$ is equal to the 83\%-light radius. Another consideration in choosing the URC was to avoid using the same velocity profile in our toy models and our method. This way, we could test the ability of our method to infer the underlying kinematics despite having different velocity profile assumptions. The URC is defined as,
\begin{equation}
    v(x) = \sqrt{v^2_d (x) + v^2_h (x)},
    \label{eq:urc_sum}
\end{equation}
where $v_d (x)$ and $ v_h (x)$ represent the disk and halo velocity component contributions with $x = r/R_\text{opt}$. The disk and halo components are defined as,
\begin{equation}
    v^2_d (x) = v^2(R_\text{opt}) \beta \frac
        {1.97 x^{1.22}}
        {(x^2 + 0.78^2)^{1.43}}
    \label{eq:urc_disc}
\end{equation}
\begin{equation}
    v^2_h (x) = 
        v^2(R_\text{opt}) (1 - \beta) (1 + \alpha^2) 
        \frac{x^2}{x^2 + \alpha^2}
    \label{eq:urc_halo}
\end{equation}
where the shape parameters are,
\begin{align}
    \alpha = 1.5 \bigg( \frac{L}{L^*} \bigg)^{1/5} &&
    \text{and} &&
    \beta = 0.72 + 0.44 \log_{10} \bigg( \frac{L}{L^*} \bigg).
    \label{eq:urc_beta}
\end{align}
We set $L/L^* = 1$ for all toy models. A systemic velocity term was omitted for simplicity. The galaxies were inclined by  45$^{\circ}$ such that the LoS velocity was observable.

The spatial edge of the cube was assumed set at 2 $R_e$. The cubes were oversampled by a factor of 5 elements in the spatial and wavelength directions. Emission lines were broadened by a Gaussian line-spread function (LSF) with $\text{FWHM}_\text{LSF} = 1.61$ \AA{} similar to the SAMI Galaxy Survey \citep{vandeSande2017} and convolved by the seeing per wavelength slice. The over-sampled data cube was integrated to the desired resolution. The resulting cubes have a $15'' \times 15''$ FoV with $30 \times 30$ elements and a wavelength range of [6554 \AA, 6571 \AA] with 31 elements. The above choices were aimed at replicating a cube cut around the H$\alpha$ emission line for a typical galaxy observed with the Sydney Australian-Astronomical-Observatory Multi-object Integral-Field Spectrograph (SAMI) instrument \citep{Croom2012}. 

To check for systematics in the kinematic inferences for different methods, we constructed the toy models with negligible noise. A grid of toy models was constructed with $\sigma_{v, \text{input}} = \{10, 20, 30, 40, 50 \} \text{ km s}^{-1}$, $v(R_\text{opt}) = \{50, 100, 150, 200, 300\} \text{ km s}^{-1}$. The toy models were convolved with a Gaussian PSF with $\text{FWHM}_\text{PSF} = \{1'', 2'', 3'' \}$ or a Moffat PSF with $\{ \text{FWHM}_\text{PSF}, \beta_\text{PSF} \} = \{ 2'', 3 \}$.

\subsubsection{Estimating the velocity dispersion}
\label{sec:systoyvdisp} % used for referring to this section from elsewhere

\begin{figure}
    \begin{center}
        \includegraphics[
            width=0.5\textwidth,
            keepaspectratio=true,
            trim=0mm 0mm 0mm 10mm,
            clip=true]{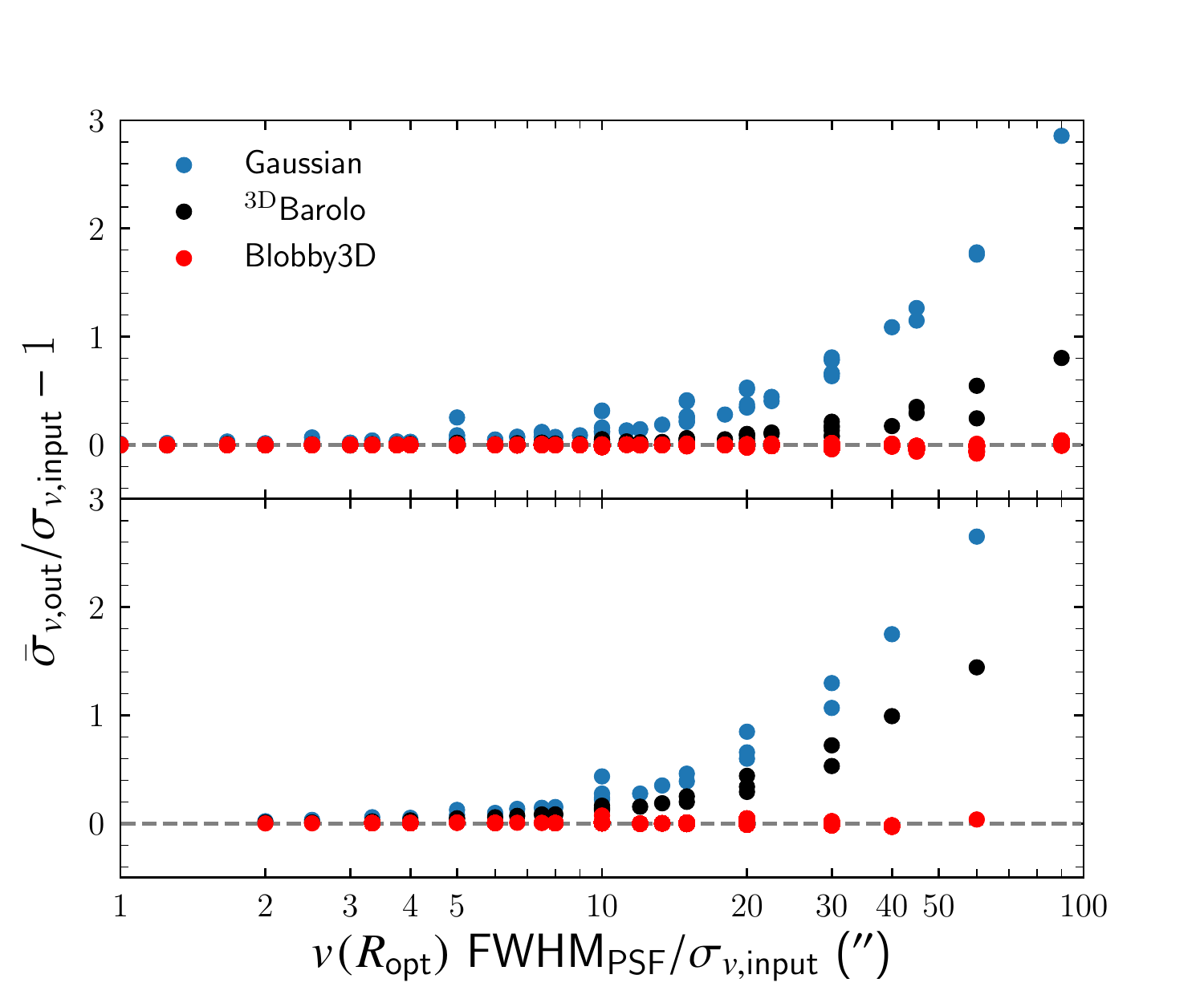}
    \end{center}
    \caption{
    Relative difference between the estimated mean velocity dispersion ($\bar{\sigma}_{v, \text{out}}$) and the input velocity dispersion ($\bar{\sigma}_{v, \text{input}}$). This is shown as a function of $v(R_\text{opt})$, the $\text{FWHM}_\text{PSF}$, and the input velocity dispersion. The methods compared were a single-component Gaussian fit to each spaxel (blue), \bbarolo{} (black), and our method (red). The model inputs are a grid of $\sigma_{v, \text{input}} = \{10, 20, 30, 40, 50\}$ km s$^{-1}$ and $v(R_{\text{opt}}) = \{50, 100, 150, 200, 300\}$ km s$^{-1}$. The PSF profiles used are a Gaussian (top) with $\text{FWHM}_\text{PSF} = \{1'', 2'', 3'' \}$ and Moffat (bottom) with $\{ \text{FWHM}_\text{PSF}, \beta_\text{PSF} \} = \{ 2'', 3 \}$. Using the mean velocity dispersion after fitting a single-component Gaussian fit per spaxel, we found that the estimated velocity dispersion increased as a function of $v(R_\text{opt}) \text{FWHM}_\text{PSF} / \sigma_{v, \text{input}}$. \bbarolo{} improves the estimates for the intrinsic mean velocity dispersion, yet still results in a trend similar to the estimates using the single-component Gaussian fit per spaxel. \blobby{} reliably infers the mean intrinsic velocity dispersion for our full grid of toy models.
    }
    \label{fig:VDispRelation}
\end{figure}

In Fig. \ref{fig:VDispRelation}, we show the relative difference between the estimated mean velocity dispersion ($\sigma_{v, \text{out}}$) and the input velocity dispersion ($\sigma_{v, \text{input}}$). The relative differences are shown compared to $v(R_\text{opt}) \text{FWHM}_\text{PSF} / \sigma_{v, \text{input}}$. This relationship yielded the clearest trend for the relative difference estimates using a single component Gaussian fit per spaxel. The intuitive reasoning for this relationship is that increasing $v(R_\text{opt}) / \sigma_{v, \text{input}}$ increases the velocity gradient at the centre of the galaxy with respect to the input velocity dispersion. This exacerbates the effect of beam smearing due to blending velocity profiles that have significantly different mean velocity compared to their width. Similarly, increasing the $\text{FWHM}_\text{PSF}$ acts to blend velocity gradients across wider regions of the galaxy.

We started by comparing a single component Gaussian fit to each each spaxel, a tilted ring model using \bbarolo, and our method. For the single-component Gaussian fits, we calculated the mean velocity dispersion of the spaxels across the FoV. The results for \bbarolo{} were calculated using the area-weighted mean velocity dispersion across the rings. For our method, we constructed the 2D velocity dispersion map for each posterior sample and then calculated the mean velocity dispersion of the spaxels. All posterior samples are shown on this plot, but due to the negligible noise applied to the toy models the posterior distributions for the mean velocity dispersion are negligible at this scale.

\begin{figure*}
    \begin{center}
    \includegraphics[
        width=0.9\textwidth,
        keepaspectratio=true,
        trim=0mm 0mm 0mm 0mm,
        clip=true]{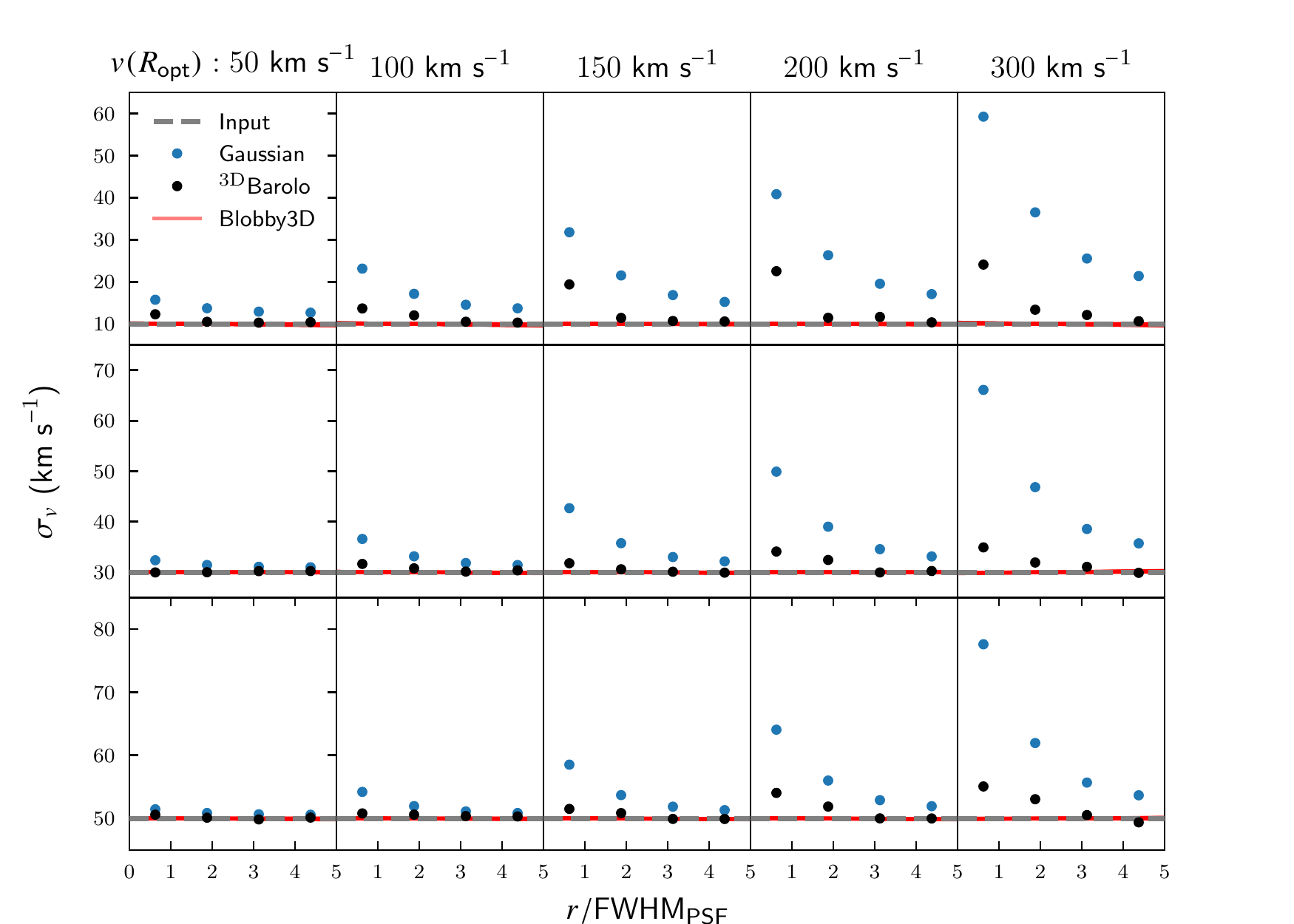}
    \end{center}
    \caption{
    Recovering the LoS intrinsic radial velocity dispersion profiles for our toy models convolved by a Guassian PSF with  $\text{FWHM}_\text{PSF} = 2''$. We show different  $v(R_{\text{opt}})$ and $\sigma_{v, \text{input}}$ per column and row, respectively. Blue points correspond to single component Gaussian fits to each spaxel and then averaged for each radial bin. Black points correspond to the velocity dispersion estimates per ring using the \bbarolo{} fitting code, and \blobby{} shows the posterior samples for the radial velocity dispersion profiles. We found that the relative difference between the estimated and actual LoS velocity dispersion increased towards the centre of the centre of the galaxy where the LoS velocity gradient is greatest. Similarly, these effects increased as $v(R_\text{opt})/\sigma_{v, \text{input}}$ increased. The estimates using \bbarolo{}  improve on the single-component Gaussian fit, while \blobby{} accurately infers the LoS velocity dispersion across the grid of toy models.
    }
    \label{fig:RadialDispToy}
\end{figure*}

To further illustrate the effect of beam smearing on the observed velocity dispersion, we show radial profiles across a grid of input $\sigma_{v, \text{input}}$ and $v(R_\text{opt})$ assuming a Gaussian convolution kernel with $\text{FWHM}_\text{PSF} = 2''$ in Fig. \ref{fig:RadialDispToy}. This shows that the effect of beam smearing increases significantly in the centre of the galaxy where the velocity gradient is highest. Increasing $v(R_\text{opt})$ also acts to increase the velocity gradient, and thus the offsets increase as well. The effect of beam smearing decreases as the input velocity dispersion increases, suggesting that the relative relationship between $v(R_\text{opt}) / \sigma_{v, \text{input}}$ is more indicative of the effects of beam smearing.

\bbarolo{} provides partial corrections for beam smearing. However, the relative difference is $\sigma_{v, \text{out}} / \sigma_{v, \text{input}} - 1 \sim 0.1$  at $v(R_\text{opt}) \text{FWHM}_\text{PSF} / \sigma_{v, \text{input}} = 30''$ and increases with $v(R_\text{opt}) \text{FWHM}_\text{PSF} / \sigma_{v, \text{input}}$. The effect of beam smearing increases towards the centre of the galaxy as well. We suspected that the observed bias was due to \bbarolo{} interpreting the unresolved velocity gradient across the discretised rings as increased velocity dispersion. Yet we found no significant difference for the estimated velocity dispersion profile when using a different number of rings. As such, the observed biases observed for \bbarolo{} appears to be fundamental for low resolution data. \citet{DiTeodoro2015} also found that \bbarolo{} over-estimated the velocity dispersion at the centre of the galaxy for low-resolution observations (see Fig. 8 in their paper).

\bbarolo{} is further affected when used for toy models convolved by a Moffat kernel. The divergence in the relative difference is $\sigma_{v, \text{out}} / \sigma_{v, \text{input}} - 1 \sim 0.1$ at $v(R_\text{opt}) \text{FWHM}_\text{PSF} / \sigma_{v, \text{input}} = 10''$. In this case, we assumed the Gaussian convolution kernel used by \bbarolo{} had a FWHM equal to that of the Moffat profile. As \bbarolo{} assumes a Gaussian PSF, we expected that using it for a toy model convolved by a Moffat kernel would affect the estimates. \citet{Bouche2015} also pointed out that significant differences for the velocity dispersion estimates can be caused by not accurately modelling the PSF axis ratio. Similar issues are likely to arise when our PSF modelling assumptions are not met. We suggest that researchers keep in mind that assumptions about the PSF will affect the velocity dispersion estimates.

\begin{figure}
    \begin{center}
        \includegraphics[
            width=0.5\textwidth,
            keepaspectratio=true,
            trim=0mm 8mm 0mm 8mm,
            clip=true]{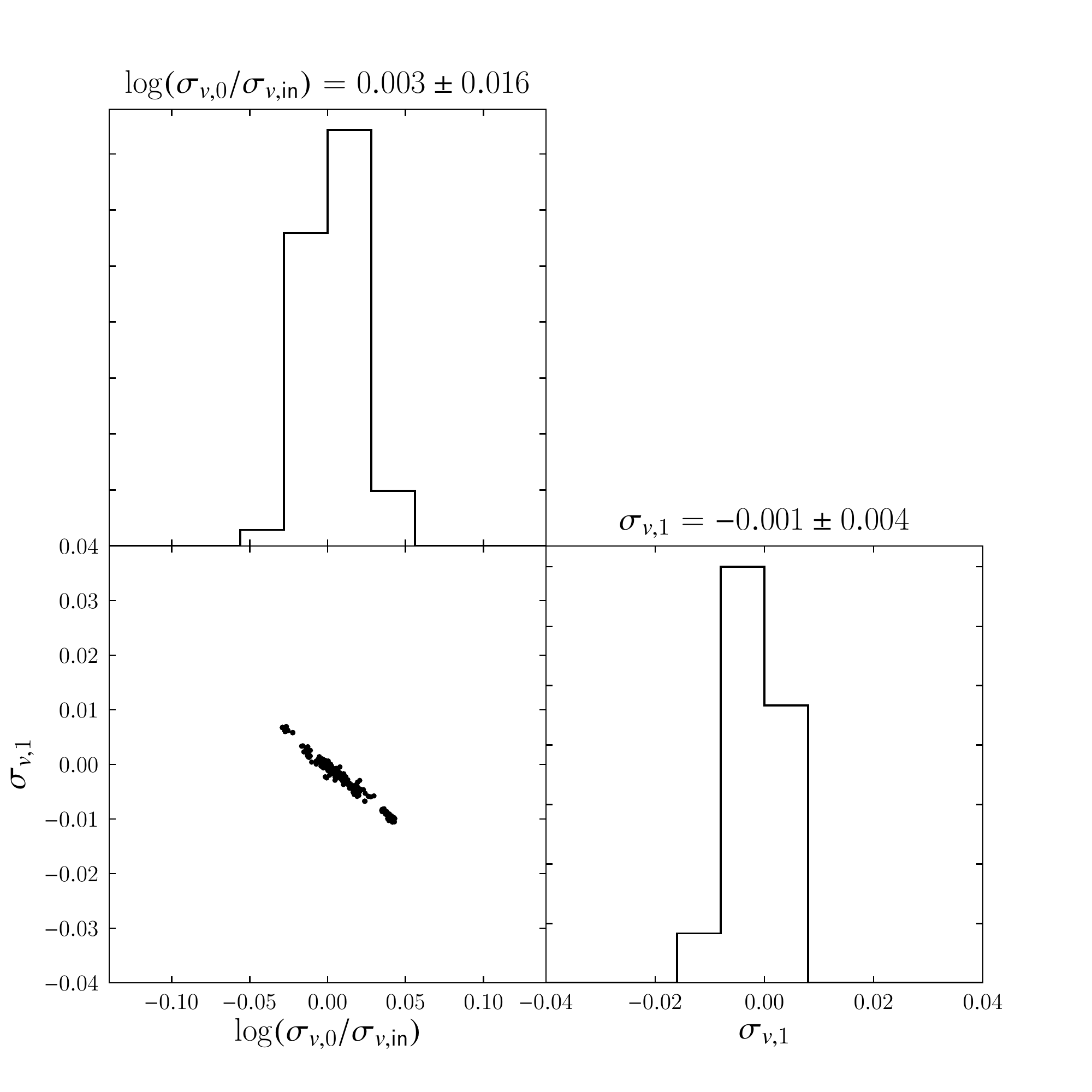}
    \end{center}
    \caption{
    Marginalised posterior distributions for the log relative difference between the modelled central velocity dispersion ($\sigma_{v, 0}$) and the input velocity dispersion ($\sigma_{v, \text{true}}$) (top), plus the log velocity dispersion gradient ($\sigma_{v, 1}$) (bottom right). We also show the conditional posterior distribution between these parameters (bottom left). We found that the distribution of our inferred intrinsic velocity dispersion parameters was consistent with our inputs.
    }
    \label{fig:CornerVDisp}
\end{figure}

Our method accurately estimates the intrinsic velocity dispersion, as shown in both the relative differences in Fig. \ref{fig:VDispRelation} and the radial profiles in Fig. \ref{fig:RadialDispToy}. We also show the posterior distribution of the log relative difference $\log({\sigma_{v, 0}} / \sigma_{v, \text{in}})$ and $\sigma_{v, 1}$ in Fig. \ref{fig:CornerVDisp}. These plots are marginalised over all toy models and the remaining parameters. The marginalised distributions remain consistent with zero for both parameters as $\log({\sigma_{V, 0}} / \sigma_{v, \text{in}}) = 0.3 \pm 1.7 \times 10^{-2}$ and $\sigma_{v, 1} = -1 \pm 4 \times 10^{-3}$. There is a slight tendency for higher $\sigma_{v, 0}$ with negative gradients, but this was negligible as the difference in velocity dispersion compared to the input values was $< 1 \text{ km s}^{-1}$ in all cases.

\subsubsection{Estimating the velocity profiles}
\label{sec:systoyvel} % used for referring to this section from elsewhere

\begin{figure*}
    \begin{center}
        \includegraphics[
            width=\textwidth,
            keepaspectratio=true,
            trim=0mm 0mm 0mm 0mm,
            clip=true]{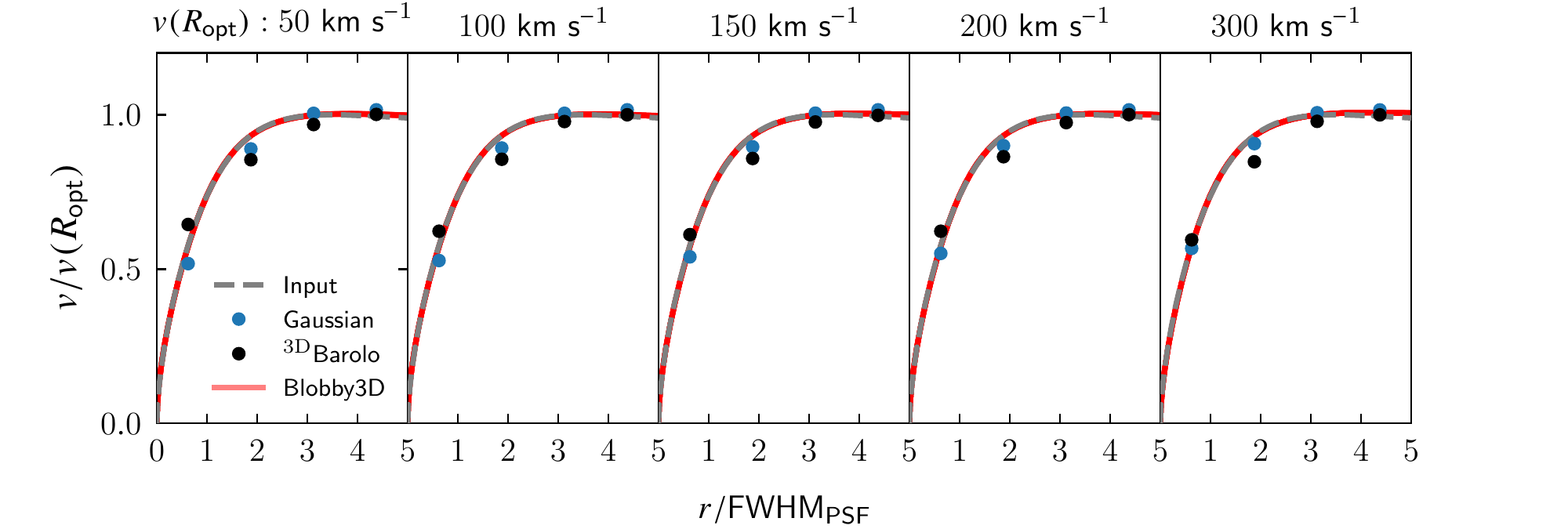}
    \end{center}
    \caption{
    Recovering the velocity profile for our toy models with exponential flux distribution, universal rotation curve with different $v(R_{\text{opt}})$, and $\sigma_v = 20 \text{ km s}^{-1}$. The toy models were convolved with a Gaussian profile with $\text{FWHM}_\text{PSF} = 2''$. The toy models were constructed with negligible noise to check for systematic biases in the methodologies. Blue dots correspond to a single component Gaussian fit to each spaxel where the mean has been calculated for 4 equally space bins. \bbarolo{} (black) show the radial velocity in each radial bin. Blobby3D (red) shows 12 posterior samples for the velocity profile, although the difference for each posterior sample is negligible due to zero noise applied to the toy models. \bbarolo{} does not fully recover the velocity profile at $v(R_\text{opt}) = 50$ km s$^{-1}$.
    }
    \label{fig:TMVVRopt}
\end{figure*}

\begin{figure}
    \begin{center}
    \includegraphics[
        width=0.5\textwidth,
        keepaspectratio=true,
        trim=0mm 0mm 0mm 0mm,
        clip=true]{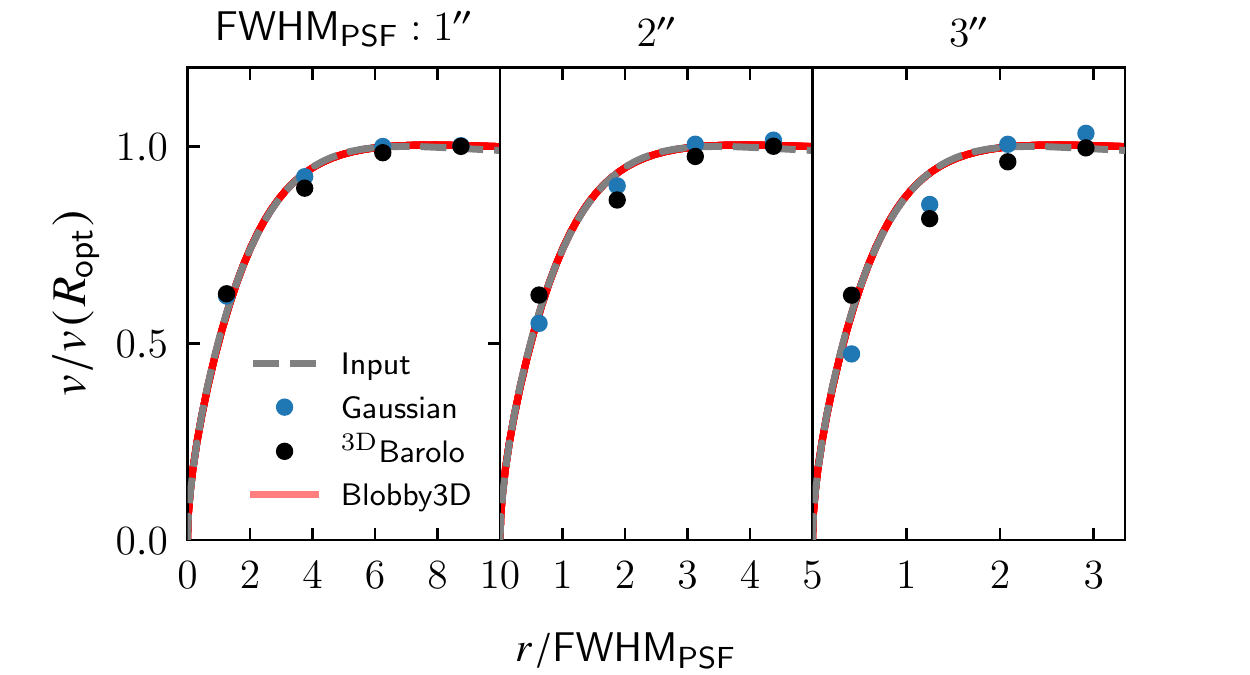}
    \end{center}
    \caption{
    Similar to Fig. \ref{fig:TMVVRopt} but setting $v(R_{\text{opt}}) = 200  \text{ km s}^{-1}$ and varying the $\text{FWHM}_\text{PSF}$. In this case, we found that the inferred LoS velocity gradient is flattened for both the single component Gaussian fits to each spaxel and \bbarolo{} as $\text{FWHM}_\text{PSF}$ increases. \blobby{} is not affected by increasing the $\text{FWHM}_\text{PSF}$.
    }
    \label{fig:TMVFWHM}
\end{figure}

We show the inferred velocity profiles for varying $v(R_\text{opt})$ and $\text{FWHM}_\text{PSF}$ in Fig.  \ref{fig:TMVVRopt} and Fig.  \ref{fig:TMVFWHM} respectively. We only show the velocity profiles for $\sigma_{v, \text{in}} = 20 \text{ km s}^{-1}$ as we did not observe any dependency on the inferred velocity profiles as a function of the input velocity dispersion.

Once again, considering the Gaussian fits as indicative for the effects of beam smearing, we note that the velocity is typically under-estimated in regions of high velocity gradient. This relative effect on the observed velocity compared to $v(R_\text{opt})$ is approximately constant. Instead, the differences are greatly affected by increasing the $\text{FWHM}_\text{PSF}$. These effects are consistent with the modelling performed by \citet{Davies2011}.

The effects of beam smearing remain when using \bbarolo{}. We did not find any significant difference for the inferred velocity profiles when we changed the number of rings.

Our method typically estimates the velocity profile well for $v(R_\text{opt}) \geq 150 \text{ km s}^{-1}$. For $v(R_\text{opt}) < 150 \text{ km s}^{-1}$, there are issues estimating the shape of the velocity profile particular in the centre of the galaxy and the outskirts. The effects for $v(R_\text{opt}) = 100 \text{ km s}^{-1}$ are minimal both in relative and absolute terms. For $v(R_\text{opt}) = 50 \text{ km s}^{-1}$ the relative difference is $\sim 0.05$ corresponding to a few km s$^{-1}$. 

The reasoning for the difference at low $v(R_\text{opt})$ remains unclear as better 1D fits for the \citet{Courteau1997} empirical model to the input Universal Rotation Curve are within the prior distribution. We suspect that the differences are driven by performing the full 3D modelling where the differences in model parameterisation and integration are slightly different for the toy modelling compared to the \blobby{} approach. However, given the negligible difference compared to systematic and variance that will be involved in modelling real data, we do not consider this to be a significant issue.

\subsection{A toy model with gas substructure}
\label{sec:BlobbyToyModel}

We then constructed a more realistic toy model. First, we constructed a toy model as defined above with $\sigma_v$ = 20 km s$^{-1}$ and $v_c$ = 200 km s$^{-1}$. We rotated the position angle of the disk by $\pi/4$ and added 10 Gaussian blobs to the gas distribution. All blobs were defined to be circular in the plane of the disk. The integrated flux for each blob was set to 10\% of the disk flux. The width for each blob was set to $w = 0.2 R_e$. The centre of the blobs were randomised uniformly with distance to the centre as $r/R_e = [0, 2]$ in the plane of the disk. We distributed the polar angle uniformly in the range $\phi_c = [0, \pi]$. We add independent and identically distributed ($iid$) Gaussian noise corresponding to mean S/N = 20 per wavelength bin. The cube was oversampled then convolved as per all of our previous toy models.

The distribution of $\phi$ in the range $[0, \pi]$ introduces an asymmetry in the flux profile as blobs are only placed on one side of the disk. We do this to show that our method is capable of recovering asymmetric gas distributions. We also note that such substructures are common in real observations.

\begin{figure*}
    \begin{center}
        \includegraphics[
            width=0.88\textwidth,
            keepaspectratio=true,
            trim=0mm 3mm 0mm 5mm,
            clip=true]{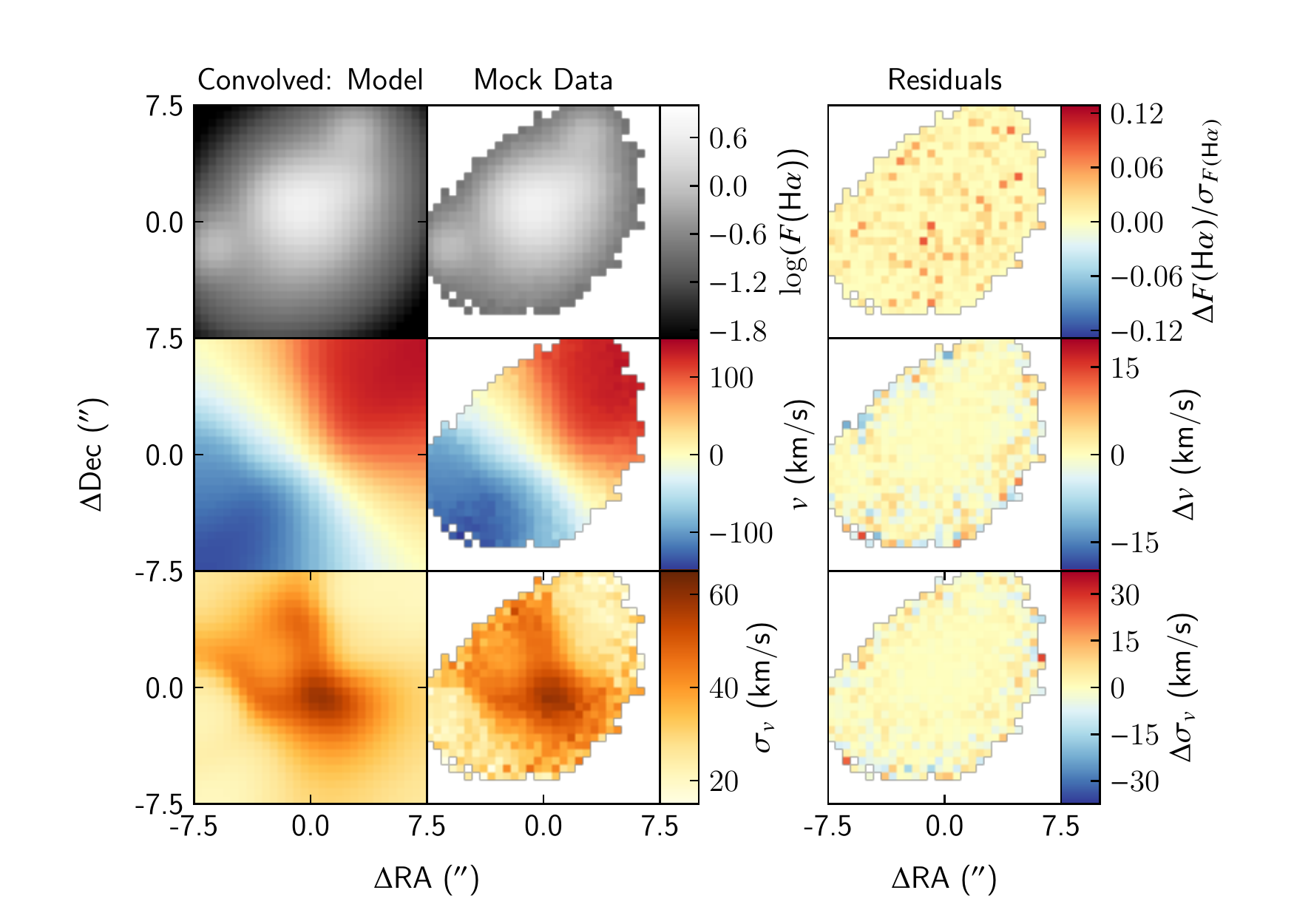} \\
        \includegraphics[
            width=0.88\textwidth,
            keepaspectratio=true,
            trim=0mm 3mm 0mm 5mm,
            clip=true]{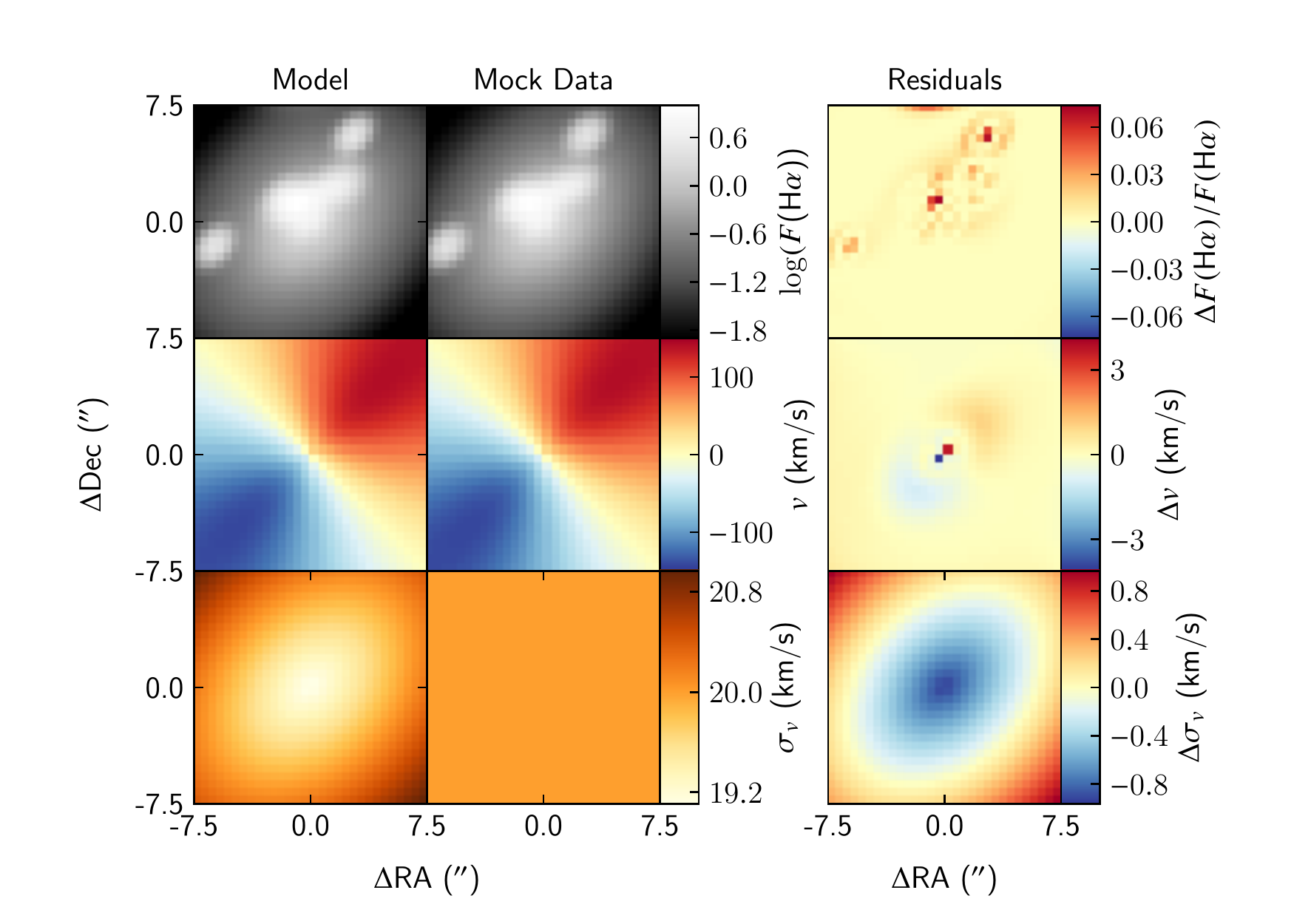}
    \end{center}
    \caption{
    2D maps for a posterior sample for a toy model with asymmetric gas substructure. For the top three rows, we show the convolved model compared to the convolved mock data (ie. toy model). The preconvolved \blobby{} model and preconvolved mock data are compared in the bottom three rows. In both cases the rows show the H$\alpha$ flux, LoS velocity profile, and LoS velocity dispersion. The columns show the respective \blobby{} output, data, and residuals. The absolute residuals are shown for the velocity and velocity dispersion maps. In the top panel we show the flux map residuals normalised with respect to the modelled Gaussian noise, whereas in the bottom panel we show the relative flux difference. The convolved mock data is shown where H$\alpha$ flux S/N > 10. We found that convolving a model with gas substructure and radial kinematic profiles introduced kinematic asymmetries. \blobby{} was able to model the gas and kinematic profile asymmetries and recover the intrinsic gas kinematics accurately.
    }
    \label{fig:blobby_sample}
\end{figure*}

We show the toy model and our results in Fig. \ref{fig:blobby_sample}. An interesting consequence of introducing asymmetries in the flux profile is that convolving the model by the PSF introduces asymmetries in the velocity dispersion profile. In this case, the 2D velocity dispersion map for the convolved data shows two tails on the side where the blobs are located.

Modelling to the convolved data is performed well with no outlying structure remaining in the residual maps. Recovery of the preconvolved model is also performed reasonably well. The maximum relative difference in the map is $\sim 0.1$ whereas the velocity profile is within several km s$^{-1}$ and the maximum difference in velocity dispersion is less than 1 km s$^{-1}$. While this posterior sample shows a very shallow positive velocity dispersion gradient (< 1 km s$^{-1}$ difference across the FoV), there is no observed bias in the gradients in the full marginalised posterior distribution with $\sigma_{v, 0} = 0.03 \pm 0.11$.

\section{Applications to Real Data}
\label{sec:SAMIModelling}

We then applied the method to a sample of 20 galaxies from the SAMI Galaxy Survey. The SAMI Galaxy Survey uses SAMI \citep{Croom2012}. SAMI uses 13 fibre bundles known as hexabundles which consist of 61 fibres with 75\% filling factor that subtend ~1.6$''$ for a total FoV with width ~15$''$ \citep{Bland-Hawthorn2011,Bryant2014}. The IFUs, as well as 26 sky fibres, are plugged into pre-drilled plates using magnetic connectors. SAMI fibres are fed to the double-beam AAOmega spectrograph \citep{Sharp2006}. The SAMI Galaxy survey uses the 570V grating at 3700-5700 \AA{} giving a resolution of $R \sim 1730$, and the R1000 grating from 6250-7350 \AA{} giving a resolution of $R \sim 4500$.

\subsection{Sample selection}
\label{sec:samisample}

The SAMI Galaxy Survey has observed > 3,000 galaxies. We aim to present initial results for a small sample of galaxies that are representative of typical star-forming galaxies within the parent sample. Star-forming galaxies were chosen as their gas kinematics typically have smoothly varying kinematic profiles. This is in contrast to galaxies with H$\alpha$ emission associated non-starforming mechanisms. A common example are galaxies with an Active Galactic Nuclei (AGN), as they typically have significantly higher velocity dispersion in the centre of the galaxy compared to the outskirts.

Star-forming galaxies were selected by applying a cutoff integrated H$\alpha$ equivalent width > 3 \AA. The equivalent width cutoff is consistent with the star-forming main sequence cutoff applied by \citet{CidFernandes2011} using single-fibre SDSS data. The equivalent width was measured as the width in the spectral dimension of a rectangle with width and height equal to a measure of the integrated continuum and H$\alpha$ flux, respectively. We used the mean continuum across the wavelength range [6500 \AA, 6540 \AA] as the estimate for the continuum per spaxel. 

We removed galaxies with H$\alpha$ emission contaminated by Active Galactic Nuclei (AGN) or LINERs using the AGN classification proposed by \citet{Kauffmann2003}. Under this classifcation, we removed galaxies under the condition that,
\begin{equation}
    \log([\text{OIII}] / H\beta) 
        > 0.61/(\log([\text{NII}]/\text{H}\alpha) - 0.05) + 1.3,
    \label{eq:kauffmanagn}
\end{equation}
where [OIII] and [NII] represent the emission lines at 5007 \AA{} and 6583 \AA{}, respectively. For each emission line, we used the integrated flux estimates in the 1.4$''$ aperture spectra data provided in the SAMI Galaxy Survey DR2 \citep{Scott2018}. The 1.4$''$ aperture spectra data are the innermost aperture spectra data provided in SAMI Galaxy Survey DR2, and thus should be the most appropriate to find galaxies with AGN or LINER emission which is typically centrally concentrated.

We selected galaxies with an intermediate inclination angle ($i \in [30^{\circ}, 60^{\circ}]$). Galaxies with low inclination were avoided as it is difficult to infer the velocity profile. Whereas galaxies close to edge-on will be difficult to model as our method assumes a thin-disk. Furthermore, galaxies observed close to edge-on are typically optically thick, such that the entire disk cannot be observed. The inclination estimates were calculated by converting an estimate for the observed ellipticity assuming a thin-disk. Similarly, we selected galaxies with intermediate effective radius ($R_e \in [2.5'', 22.5'']$). This avoids small galaxies that are not well resolved. It also ignores large galaxies which may be difficult to infer their velocity profile. Estimates for the ellipticity and effective radius were taken from the SAMI Galaxy Survey parent catalogue \citep{Bryant2015}, who in turn used the single S\'ersic fits to the $r$-band Sloan Digital Sky Survey images by \citet{Kelvin2012}.

There are 330 galaxies that meet the above criteria in the SAMI Galaxy Survey DR2. We chose 20 galaxies with our final galaxy sample shown in Table \ref{tab:samisample}.

\begin{table*}
    \caption{
    Summary statistics for our sample of galaxies from the SAMI galaxy survey. All values are sourced from the SAMI parent catalogue described by \citet{Bryant2015}. We also show the estimated SAMI Galaxy Survey pipeline estimated values for the PSF assuming a Moffat profile.
    }
    \begin{center}
        \begin{tabular}{l ccc ccccc}
            \hline
            GAMA ID & 
            RA & 
            Dec & 
            $z_{\text{spec}}$ & 
            $\log( M_{*}/M_\odot )$ & 
            $R_e$ & 
            $e$ & 
            FWHM$_{\text{PSF}}$ &
            $\beta_{\text{PSF}}$ 
            \\
            & 
            $(^{\circ})$ & 
            $(^{\circ})$ & 
            & 
            & 
            ($''$) & 
            & 
            ($''$) &
            \\
            \hline
            214245 & 129.52446 & 0.60896 & 0.014 & 9.40 & 4.46 & 0.32 & 2.12 & 3.65 \\
            220371 & 181.23715 & 1.50824 & 0.020 & 9.53 & 6.97 & 0.35 & 3.37 & 6.78 \\
            220578 & 182.17817 & 1.45636 & 0.019 & 8.98 & 2.96 & 0.41 & 2.34 & 2.71 \\
            238395 & 214.24319 & 1.64043 & 0.025 & 9.87 & 4.11 & 0.18 & 3.29 & 4.76 \\
            273951 & 185.93037 & 1.31109 & 0.026 & 8.72 & 4.34 & 0.45 & 1.62 & 2.77 \\
            278804 & 133.85939 & 0.85818 & 0.042 & 9.82 & 2.65 & 0.38 & 2.87 & 4.03 \\
            298114 & 218.40091 & 1.30590 & 0.056 & 10.25 & 4.84 & 0.41 & 2.26 & 4.01 \\
            30346 & 174.63865 & -1.18449 & 0.021 & 10.45 & 11.25 & 0.32 & 1.89 & 2.48 \\
            30377 & 174.82286 & -1.07931 & 0.027 & 8.22 & 3.81 & 0.35 & 3.30 & 3.81 \\
            30890 & 177.25796 & -1.10260 & 0.020 & 9.79 & 7.56 & 0.43 & 2.92 & 3.94 \\
            422366 & 130.59560 & 2.49733 & 0.029 & 9.62 & 8.86 & 0.49 & 1.78 & 2.49 \\
            485885 & 217.75790 & -1.71721 & 0.055 & 10.25 & 5.04 & 0.16 & 2.27 & 5.19 \\
            517167 & 131.16137 & 2.41098 & 0.030 & 9.24 & 3.67 & 0.31 & 2.01 & 2.81 \\
            55367 & 181.79334 & -0.25959 & 0.022 & 8.40 & 6.71 & 0.30 & 1.56 & 3.64 \\
            56183 & 184.85245 & -0.29410 & 0.039 & 9.50 & 3.58 & 0.23 & 2.18 & 3.19 \\
            592999 & 215.06156 & -0.07938 & 0.053 & 10.26 & 4.24 & 0.47 & 1.53 & 2.98 \\
            617655 & 212.63506 & 0.22418 & 0.029 & 9.07 & 5.08 & 0.14 & 2.85 & 8.67 \\
            69620 & 175.72473 & 0.16189 & 0.018 & 9.30 & 4.45 & 0.25 & 2.53 & 4.49 \\
            84107 & 175.99843 & 0.42801 & 0.029 & 9.71 & 5.05 & 0.23 & 2.53 & 4.49 \\
            85423 & 182.27832 & 0.47328 & 0.020 & 8.63 & 3.56 & 0.18 & 2.90 & 3.55 \\
            \hline
        \end{tabular}
    \end{center}
    \label{tab:samisample}
\end{table*}

\subsection{Data cubes}
\label{sec:samicubes}

The data cubes we used were from the SAMI internal data release v0.10.1 \citep{Scott2018}. Data cubes were redshift corrected by the spectroscopic redshift which was taken from the SAMI parent catalogue \citep{Bryant2015} who used the estimates from the Galaxy and Mass Assembly (GAMA) survey \citep{Driver2011}.

The data cubes were then cut around the H$\alpha$ emission line by $\pm 500 \text{ km s}^{-1}$. In our sample, this was wide enough to observe the full H$\alpha$ emission line while avoiding significant influence from the adjacent [NII] emission lines.

The continuum model used to subtract from the data cubes were the single-component \lzifu{} \citep{Ho2016LZIFU} data products from the SAMI Galaxy Survey internal data release v0.10.1. \lzifu{} uses the penalised pixel-fitting routine \citep[\textsc{pPXF},][]{Cappellari2004} to model the continuum using a combination of spectral stellar population templates.

Poor continuum modelling can cause systematics in the data cube that are not well represented in the galaxy model parameterisation. While we could extend the systematic parameterisation to account for systematics introduced by poor continuum modelling, such corrections would likely require a large number of nuisance parameters that would be difficult to marginalise over. Instead, we masked pixels with H$\alpha$ flux signal-to-noise < 3 and performed a secondary fit to the data using a Gaussian plus linear continuum estimate to the region cut around the H$\alpha$ line. The continuum estimated from this fit was then subtracted from the data.

\subsection{Results}
\label{sec:samiresults}

\begin{figure*}
    \begin{center}
        \includegraphics[
            width=\textwidth,
            keepaspectratio=true,
            trim=8mm 0mm 0mm 10mm,
            clip=true]{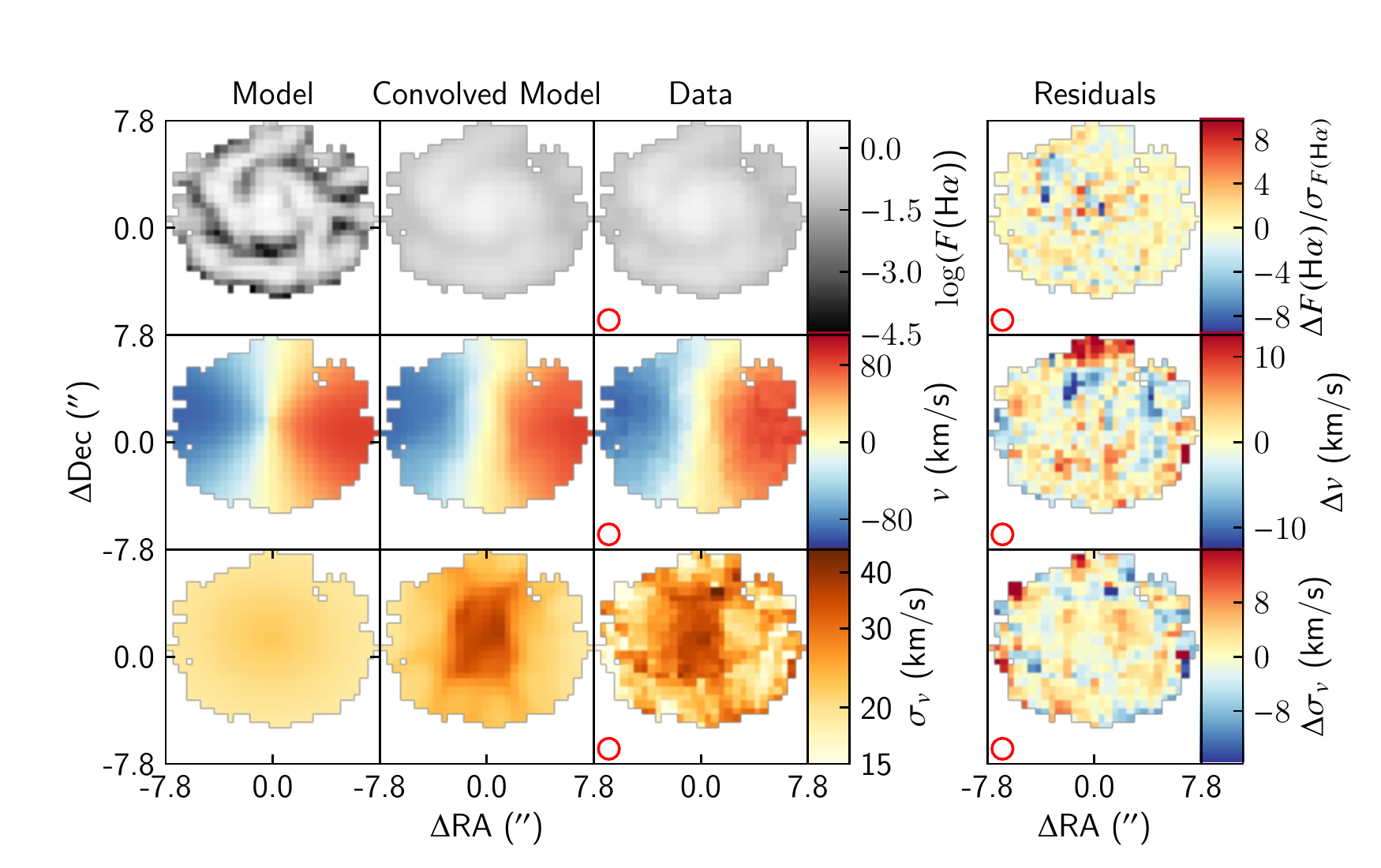} \\
        \includegraphics[
            width=\textwidth,
            keepaspectratio=true,
            trim=8mm 0mm 0mm 10mm,
            clip=true]{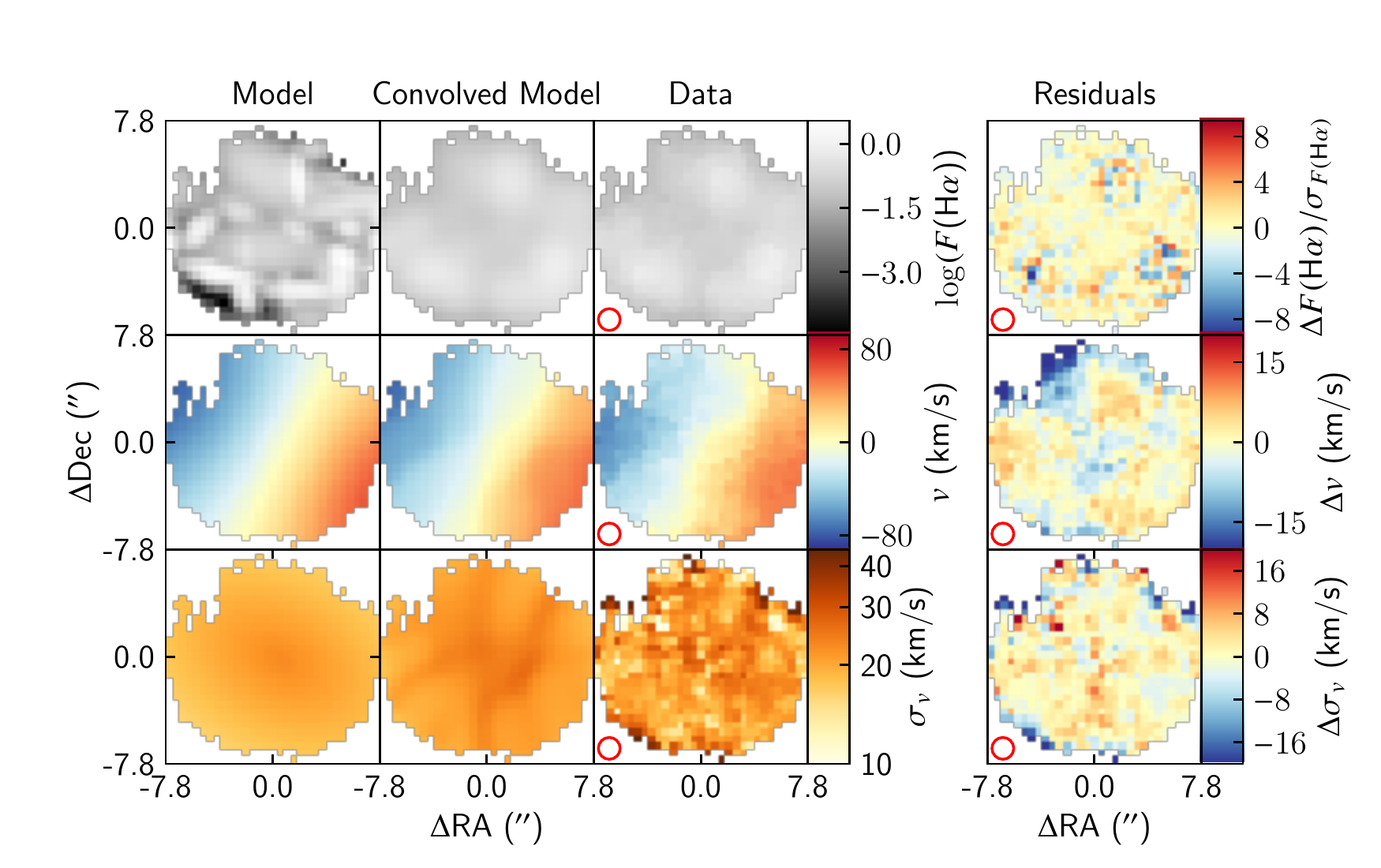}
    \end{center}
    \caption{
    2D maps for a single posterior sample for GAMA 485885 (top) and 220371 (bottom). For each galaxy we show from left to right the model, convolved model, single-component Gaussian fits to the data, and 2D residuals where $\Delta F(\text{H} \alpha) = F(\text{H} \alpha, \text{Convolved Model}) - F(\text{H} \alpha, \text{Data})$. The flux map residuals have been normalised with respect to the modelled Gaussian noise, whereas the absolute difference is shown for the velocity and velocity dispersion maps. Red circles with $r = \text{FWHM}_\text{PSF}$ indicate the seeing width. The rows show the H$\alpha$ flux, LoS velocity profile, and LoS velocity dispersion. Spaxels are shown where the data H$\alpha$ flux S/N > 10. These examples show the ability of \blobby{} to model galaxies with spirals and clumpy profiles. Parameterising complex gas distributions such as observed in the these galaxies is typically difficult, but they are a natural output of our approach.
    }
    \label{fig:example_sample1}
\end{figure*}

\begin{figure*}
    \begin{center}
        \includegraphics[
            width=\textwidth,
            keepaspectratio=true,
            trim=8mm 3mm 0mm 5mm,
            clip=true]{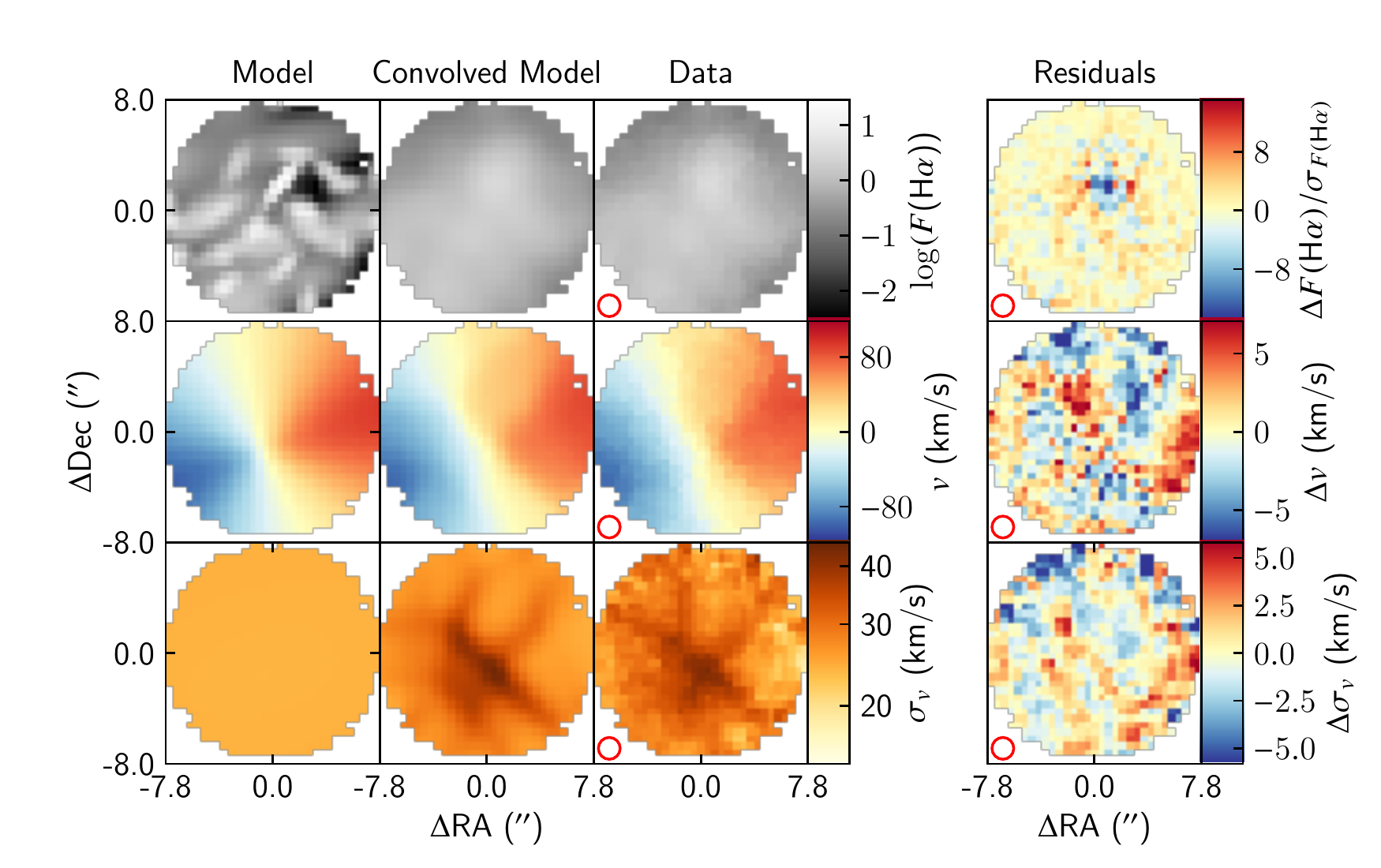}
    \end{center}
    \caption{
    Same as Fig. \ref{fig:example_sample1} for GAMA 30890. This galaxy exhibits asymmetric substructure in the H$\alpha$ gas kinematic maps. \blobby{} partially recovers the kinematic asymmetries despite only introducing asymmetric substructure in the H$\alpha$ gas distribution. This is similar to the asymmetries modelled for our toy model with gas substructure in Fig. \ref{fig:blobby_sample}. This suggests that beam smearing can play a role in the observed substructure for the H$\alpha$ gas kinematics.
    }
    \label{fig:example_sample2}
\end{figure*}

For completeness, we show our estimates of the marginalised distributions for all parameters, omitting individual blob parameters, in Tables \ref{tab:samiglobalinf} and \ref{tab:samikininf}. We also show 2D maps of an example posterior sample for GAMA 485885 and 220371 in Fig. \ref{fig:example_sample1}. A galaxy with asymmetric substructure observed in the gas kinematics is shown in \ref{fig:example_sample2}. These example posterior samples show the ability of our method to fit complex substructure. Note that the exact shape of each blob does change per posterior sample, so these should only be considered for illustrative purposes.

\begin{table*}
    \caption{
    Inferences for global parameters and blob hyperparameters for our sample of galaxies from the SAMI Galaxy Survey. We show the mean and standard deviation for the marginalised distribution for each parameter. Note that flux units are 10$^{-16}$ erg s$^{-1}$ cm$^{-2}$.
    }
    \centering
    \begin{tabular}{l rrrr rrrrr}
        \hline
        GAMA ID & 
        $N$ & 
        PA &
        $\mu_r$ &
        $\mu_F$ &
        $\sigma_F$ &
        $W_{\text{max}}$ &
        $q_{\text{min}}$ &
        $\log(\sigma_0)$
        \\
        & 
        &
        $(^{\circ})$ &
        ($''$) &
        &
        &
        ($''$) &
        &
        \\
        \hline
        214245 & 79$\pm$17 & 304.4$\pm$0.2 & 24$\pm$5 & 2.5$\pm$0.6 & 1.2$\pm$0.3 & 1.05$\pm$0.07 & 0.206$\pm$0.007 & -3.43$\pm$0.04 \\
        220371 & 117$\pm$20 & 332.05$\pm$0.09 & 25$\pm$4 & 4.2$\pm$1.0 & 1.3$\pm$0.2 & 1.81$\pm$0.06 & 0.25$\pm$0.02 & -3.09$\pm$0.01 \\
        220578 & 20$\pm$7 & 22.2$\pm$0.1 & 13$\pm$7 & 46$\pm$7 & 0.4$\pm$0.2 & 0.504$\pm$0.004 & 0.29$\pm$0.04 & -2.445$\pm$0.004 \\
        238395 & 173$\pm$45 & 163.11$\pm$0.07 & 21$\pm$6 & 20$\pm$11 & 1.9$\pm$0.4 & 0.5009$\pm$0.0008 & 0.36$\pm$0.02 & -1.569$\pm$0.002 \\
        273951 & 15$\pm$4 & 30.2$\pm$0.7 & 5$\pm$3 & 28$\pm$14 & 1.8$\pm$0.4 & 0.5005$\pm$0.0005 & 0.22$\pm$0.02 & -1.508$\pm$0.003 \\
        278804 & 18$\pm$3 & 209$\pm$1 & 1.9$\pm$0.7 & 4$\pm$1 & 1.0$\pm$0.2 & 0.51$\pm$0.02 & 0.28$\pm$0.07 & -2.07367$\pm$0.00003 \\
        298114 & 112$\pm$25 & 272.80$\pm$0.03 & 21$\pm$6 & 15.4$\pm$0.7 & 0.32$\pm$0.05 & 1.92$\pm$0.04 & 0.203$\pm$0.004 & -2.461$\pm$0.005 \\
        30346 & 70$\pm$9 & 304.32$\pm$0.02 & 26$\pm$5 & 72$\pm$5 & 0.46$\pm$0.08 & 2.31$\pm$0.04 & 0.23$\pm$0.01 & -2.009$\pm$0.003 \\
        30377 & 79$\pm$23 & 173$\pm$1 & 21$\pm$6 & 1.1$\pm$0.4 & 1.0$\pm$0.3 & 0.51$\pm$0.01 & 0.7$\pm$0.2 & -8$\pm$2 \\
        30890 & 100$\pm$17 & 19.35$\pm$0.03 & 22$\pm$5 & 23$\pm$2 & 0.71$\pm$0.08 & 2.31$\pm$0.05 & 0.24$\pm$0.01 & -2.587$\pm$0.003 \\
        422366 & 159$\pm$29 & 258.37$\pm$0.10 & 26$\pm$3 & 5$\pm$1 & 0.9$\pm$0.2 & 0.6$\pm$0.1 & 0.23$\pm$0.03 & -2.2229$\pm$0.0002 \\
        485885 & 130$\pm$33 & 353.0$\pm$0.1 & 19$\pm$6 & 2.8$\pm$0.4 & 0.79$\pm$0.09 & 0.5007$\pm$0.0008 & 0.202$\pm$0.002 & -3.38$\pm$0.03 \\
        517167 & 59$\pm$16 & 359.58$\pm$0.10 & 21$\pm$6 & 7$\pm$3 & 1.5$\pm$0.3 & 1.2$\pm$0.1 & 0.203$\pm$0.004 & -2.634$\pm$0.004 \\
        55367 & 177$\pm$46 & 182.8$\pm$0.1 & 24$\pm$4 & 0.5$\pm$0.2 & 1.5$\pm$0.2 & 0.79$\pm$0.04 & 0.202$\pm$0.002 & -8$\pm$2 \\
        56183 & 115$\pm$34 & 264.27$\pm$0.07 & 15$\pm$7 & 1.9$\pm$0.7 & 1.9$\pm$0.3 & 1.18$\pm$0.03 & 0.42$\pm$0.03 & -2.760$\pm$0.004 \\
        592999 & 98$\pm$22 & 223.90$\pm$0.05 & 20$\pm$6 & 7$\pm$1 & 1.0$\pm$0.1 & 2.24$\pm$0.05 & 0.207$\pm$0.005 & -2.362$\pm$0.003 \\
        617655 & 117$\pm$26 & 316.5$\pm$0.1 & 23$\pm$5 & 1.9$\pm$0.4 & 1.0$\pm$0.1 & 1.29$\pm$0.04 & 0.42$\pm$0.02 & -8$\pm$2 \\
        69620 & 152$\pm$25 & 300.20$\pm$0.07 & 23$\pm$4 & 17$\pm$2 & 0.65$\pm$0.06 & 0.5002$\pm$0.0002 & 0.28$\pm$0.03 & -2.072$\pm$0.002 \\
        84107 & 110$\pm$23 & 274.66$\pm$0.04 & 23$\pm$5 & 19$\pm$4 & 1.2$\pm$0.2 & 0.5001$\pm$0.0001 & 0.544$\pm$0.010 & -1.775$\pm$0.002 \\
        85423 & 87$\pm$24 & 251.2$\pm$0.3 & 23$\pm$5 & 1.1$\pm$0.7 & 1.6$\pm$0.4 & 1.09$\pm$0.05 & 0.48$\pm$0.06 & -8$\pm$2 \\
        \hline
    \end{tabular}
    \label{tab:samiglobalinf}
\end{table*}

\begin{table*}
    \caption{
    Inferences for galaxy kinematic parameters for our sample of galaxies from the SAMI Galaxy Survey.  We show the mean and standard deviation for the marginalised distribution for each parameter.
    }
    \centering
    \begin{tabular}{l rrr rrrrr rrrrr}
        \hline
        GAMA ID & 
        $v_{\text{sys}}$ & 
        $v_c$& 
        $r_t$& 
        $\gamma_v$ &
        $\beta_v$ &
        $\sigma_{v, 0}$ & 
        $\sigma_{v, 1}$ 
        \\
        & 
        ($\text{km s}^{-1}$) & 
        ($\text{km s}^{-1}$) &
        ($''$) &
        &
        &
        (km s$^{-1}$) &
        &
        \\
        \hline
        214245 & -11.5$\pm$0.1 & 71$\pm$1 & 3.69$\pm$0.04 & 81$\pm$12 & -0.36$\pm$0.03 & 25.7$\pm$0.6 & -0.087$\pm$0.005 \\
        220371 & -5.03$\pm$0.08 & 178$\pm$5 & 8.0$\pm$0.3 & 1.43$\pm$0.08 & -0.24$\pm$0.03 & 23.0$\pm$0.5 & -0.031$\pm$0.003 \\
        220578 & -15.6$\pm$0.3 & 72$\pm$1 & 6.2$\pm$0.1 & 58$\pm$22 & 0.71$\pm$0.01 & 20.6$\pm$0.5 & -0.104$\pm$0.009 \\
        238395 & -3.58$\pm$0.08 & 147$\pm$3 & 2.5$\pm$0.4 & 1.03$\pm$0.03 & 0.31$\pm$0.05 & 27.3$\pm$0.2 & 0.023$\pm$0.002 \\
        273951 & 5.95$\pm$0.08 & 242$\pm$81 & 15$\pm$3 & 17$\pm$24 & -0.71$\pm$0.04 & 33.0$\pm$0.9 & -0.17$\pm$0.03 \\
        278804 & -16.4$\pm$0.8 & 140$\pm$2 & 6.5$\pm$0.2 & 3.8$\pm$0.5 & 0.662$\pm$0.008 & 26$\pm$2 & -0.21$\pm$0.05 \\
        298114 & 5.19$\pm$0.07 & 180.6$\pm$0.4 & 2.051$\pm$0.009 & 93$\pm$8 & -0.149$\pm$0.006 & 21.4$\pm$0.3 & 0.001$\pm$0.002 \\
        30346 & 2.09$\pm$0.08 & 183.7$\pm$0.2 & 0.684$\pm$0.009 & 94$\pm$6 & -0.08$\pm$0.02 & 12.3$\pm$0.3 & 0.051$\pm$0.003 \\
        30377 & 5.4$\pm$0.3 & 274$\pm$55 & 13$\pm$2 & 21$\pm$24 & -0.70$\pm$0.04 & 18.1$\pm$0.5 & 0.023$\pm$0.006 \\
        30890 & -7.64$\pm$0.05 & 134.0$\pm$0.4 & 1.20$\pm$0.07 & 1.24$\pm$0.02 & -0.47$\pm$0.06 & 23.7$\pm$0.1 & 0.001$\pm$0.001 \\
        422366 & -12.8$\pm$0.3 & 78$\pm$1 & 5.53$\pm$0.07 & 17$\pm$8 & 0.52$\pm$0.02 & 18.3$\pm$0.4 & 0.018$\pm$0.003 \\
        485885 & -5.6$\pm$0.1 & 129$\pm$6 & 4.3$\pm$0.1 & 2.8$\pm$0.4 & 0.67$\pm$0.03 & 21.8$\pm$0.3 & -0.017$\pm$0.003 \\
        517167 & -9.80$\pm$0.10 & 73.5$\pm$0.5 & 4.38$\pm$0.02 & 92$\pm$8 & 0.593$\pm$0.009 & 13.8$\pm$0.2 & 0.075$\pm$0.004 \\
        55367 & -10.2$\pm$0.1 & 70$\pm$5 & 27$\pm$2 & 32$\pm$26 & 0.40$\pm$0.02 & 14.9$\pm$0.7 & -0.15$\pm$0.01 \\
        56183 & -6.99$\pm$0.09 & 111$\pm$2 & 4.8$\pm$0.3 & 1.21$\pm$0.03 & 0.58$\pm$0.01 & 31.6$\pm$0.2 & -0.076$\pm$0.002 \\
        592999 & -17.00$\pm$0.10 & 185$\pm$3 & 7.5$\pm$0.1 & 1.85$\pm$0.06 & 0.601$\pm$0.009 & 33.8$\pm$0.5 & -0.061$\pm$0.003 \\
        617655 & 8.07$\pm$0.09 & 86$\pm$3 & 3.9$\pm$0.1 & 10$\pm$3 & 0.26$\pm$0.06 & 14.0$\pm$0.4 & 0.039$\pm$0.006 \\
        69620 & 3.86$\pm$0.08 & 106$\pm$3 & 19.5$\pm$0.9 & 2.7$\pm$0.2 & 0.590$\pm$0.008 & 20.6$\pm$0.1 & 0.021$\pm$0.001 \\
        84107 & 7.25$\pm$0.06 & 99.5$\pm$0.6 & 3.62$\pm$0.01 & 96$\pm$4 & 0.34$\pm$0.01 & 25.1$\pm$0.3 & 0.017$\pm$0.003 \\
        85423 & 94$\pm$2 & 177$\pm$5 & 5.33$\pm$0.09 & 32$\pm$22 & -0.68$\pm$0.03 & 19$\pm$1 & -0.07$\pm$0.02 \\
        \hline
    \end{tabular}
    \label{tab:samikininf}
\end{table*}

For GAMA 485885, we see the ability of our approach to resolve a classic spiral gas distribution. The 2D residuals between the convolved model and data exhibit significant differences on scales less than the $\text{FWHM}_\text{PSF}$. The 2D maps for the LoS gas kinematics suggest that the gas is approximately regularly rotating around a kinematic centre, potentially with a small warp in the kinematic position angle. The H$\alpha$ gas velocity dispersion is peaked within the centre of the galaxy as expected for most regularly rotating galaxies that have been affected by beam smearing.

We show 2D maps for GAMA 220371 in Fig. \ref{fig:example_sample1}. This galaxy has a clumpy H$\alpha$ gas profile. We are still able to construct an adequate model to the data using our approach. The 2D residual maps for the H$\alpha$ flux show greater differences for three clumps in the North-East, South-East, and South-West regions. However, the general structure of the clumps is reasonably well resolved. The maps for the gas kinematics suggest an approximately regularly rotating galaxy. The velocity dispersion map does not show a significant peak in the centre of the galaxy compared to GAMA 485885. This is likely driven by having a shallower LoS velocity gradient and less centralised H$\alpha$ gas flux compared to GAMA 485885.

An example posterior sample for GAMA 30890 is shown in Fig. \ref{fig:example_sample2}. This galaxy exhibits asymmetries and substructure in the LoS H$\alpha$ gas kinematic maps. We are able to partially recover the H$\alpha$ gas kinematics despite only introducing asymetries in the gas H$\alpha$ gas distribution. Some substructure in the residuals remain with a patch of H$\alpha$ gas flux that is lower in the convolved model compared to data. There is also a slight warp in the LoS velocity profile as a function of radius, and differences in the velocity dispersion on the order of 5 km s$^{-1}$. However, the convolved model still performs reasonably well at resolving the gas flux and kinematics. Our ability to partially resolve the gas kinematic asymmetries, suggests that the H$\alpha$ gas distribution plus beam smearing can result in gas substructures that are not necessarily present in the underlying data. This is similar to the results we saw in Fig. \ref{fig:blobby_sample}, where we showed that introducing asymmetric substructure in the gas distribution for a regularly rotating toy model plus beam smearing led to substructure in the gas kinematics.

The SAMI Galaxy Survey provides gas kinematic data products estimated using the \lzifu{} package \citep{Ho2016LZIFU}. \lzifu{} performs single and multiple Gaussian component fits to the emission lines. Corrections for instrumental broadening are performed by subtracting the LSF from the velocity dispersion in quadrature. Effects of beam smearing are not considered.

\begin{figure}
    \begin{center}
        \includegraphics[
            width=0.5\textwidth,
            keepaspectratio=true,
            trim=2mm 2mm 10mm 5mm,
            clip=true
            ]{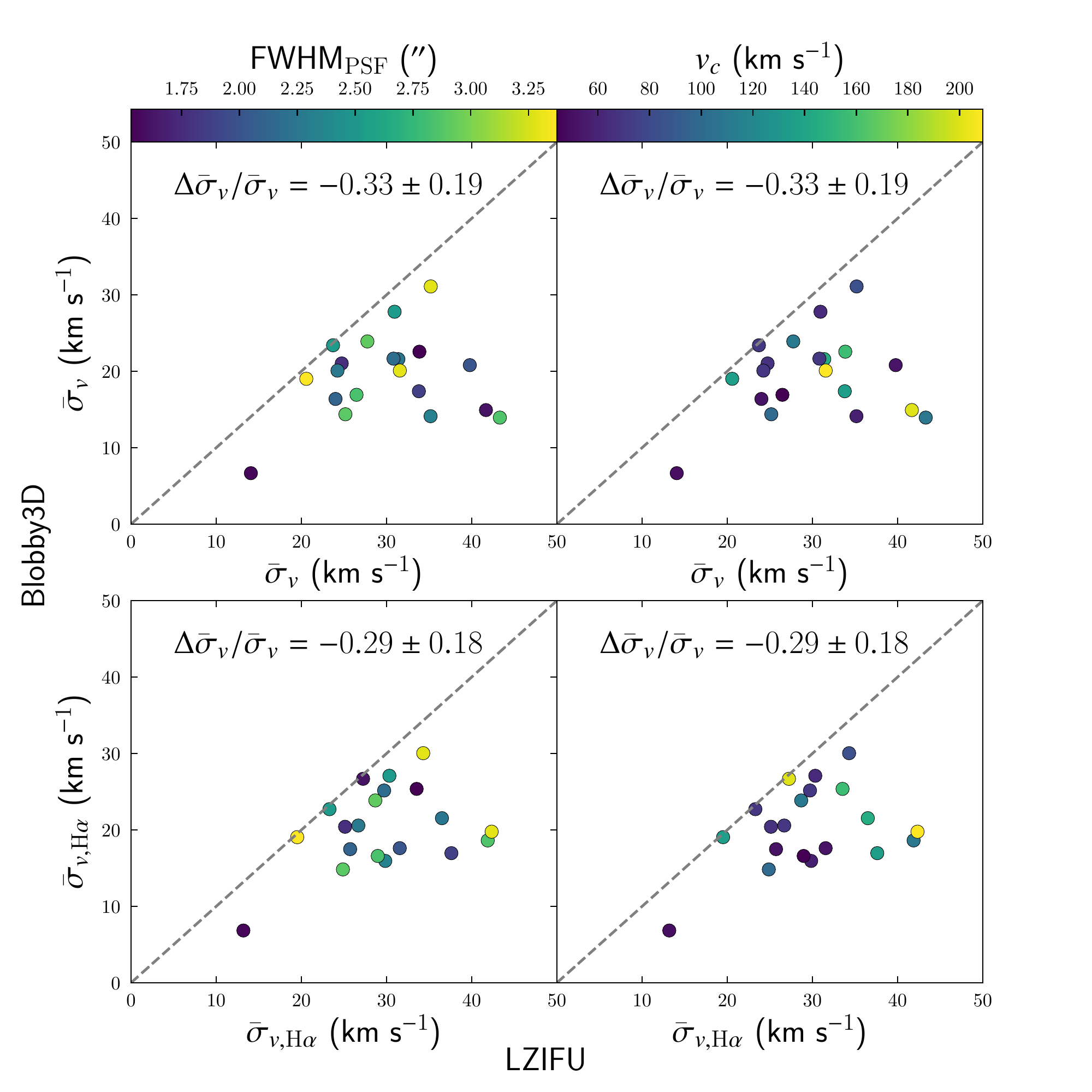}
    \end{center}
    \caption{
    Comparing estimates for the mean velocity dispersion using maps from the \lzifu{} data products and 2D maps of our method. The comparisons calculated for the unweighted (top) and H$\alpha$ flux-weighted (bottom) mean of the 2D velocity dispersion maps. $\Delta \bar{\sigma}_v / \bar{\sigma}_v$ is the arithmetic mean relative correction. We found that \blobby{} made significant corrections to the velocity dispersion estimates inferred by \lzifu.
    }
    \label{fig:blobbyvlzifu}
\end{figure}

A comparison between inferences for the global velocity dispersion between the single component \lzifu{} data products and our method are shown in Fig. \ref{fig:blobbyvlzifu}. We compare the uniformly weighted ($\bar\sigma_v$) and H$\alpha$ flux-weighted ($\bar\sigma_{v, \text{H}\alpha}$) mean velocity dispersion across the FoV. We only consider spaxels with H$\alpha$ signal-to-noise > 10 as estimated by \lzifu. This was primarily due to the increased scatter in the \lzifu{} estimates for the H$\alpha$ gas velocity dispersion in the low signal-to-noise regions.

Estimates of the global velocity dispersion using our method are in the range $\sim$[7, 30] km s$^{-1}$ using both the unweighted and H$\alpha$ flux-weighted mean. This is in comparison to estimates using the single component \lzifu{} data products of $\sim [10, 45]$ km s$^{-1}$. 

The mean relative corrections per galaxy ($\Delta \bar{\sigma}_v / \bar{\sigma}_V$) from our method is $-0.33 \pm 0.19 $ and $-0.29 \pm 0.18$ when comparing $\bar\sigma_v$ and $\bar\sigma_{v, \text{H}\alpha}$, respectively. Absolute corrections for the H$\alpha$ flux-weighted mean velocity dispersion were 
$-9^{+7}_{-13}$ km s$^{-1}$.

In Fig. \ref{fig:blobbyvlzifu}, the data are colour-coded by $\text{FWHM}_\text{PSF}$ (left) and $v_c$ (right). Qualitatively, we do not find significant trends for our corrections as a function of either of these parameters. We did expect to see a relationship between these parameters and our velocity dispersion corrections as that would be consistent with our toy model results. A larger sample of galaxies is probably required to find clear relationships between these variables and our corrections.

\section{Discussion}
\label{sec:Discussion}

\subsection{Estimating global velocity dispersion}
\label{sec:GlobalVDisp}

Beam smearing is well-known to researchers that study spatially resolved spectroscopy. As such, there have been a number of approaches to correct for beam smearing in the literature. Most of this focus has been on correcting for the global velocity dispersion. 

\subsubsection{Heuristic Approaches}
\label{sec:Heuristic}

A number of heuristic calculational approaches have been developed in an effort to estimate the intrinsic global velocity dispersion. A popular approach is to calculate an estimator of the velocity dispersion in regions away from the centre of the galaxy where beam smearing is expected to be negligible \citep[eg.][]{Johnson2018}.

Another approach is to perform corrections for a global velocity dispersion estimator as a function of factors that drive beam smearing. For example, \citet{Johnson2018} derived corrections for the median velocity dispersion and the velocity dispersion in the outskirts of the galaxy as a function of the rotational velocity and the PSF width compared to the disk width. The functional form was estimated using a grid of toy models. Using this method, they estimated relative corrections for the median velocity dispersion as $\Delta \bar{\sigma}_v / \bar{\sigma}_v = 0.2^{+0.3}_{-0.1}$ for a sample of star-forming galaxies at $z \sim 1$ using data from the KMOS Redshift One Spectroscopic Survey (KROSS). Their relative corrections for the velocity dispersion are similar to those found in this paper. However, the median seeing for KROSS was 0.7$''$ corresponding to 5.4 kpc at the median redshift of their sample. In comparison, the mean seeing for the SAMI Galaxy Survey is 2.06$''$ \citep{Scott2018}, corresponding to 1.75 kpc at the mean redshift of $z = 0.043$ of the full SAMI Galaxy Survey sample. As such, the effect due to beam smearing on the observed velocity dispersion are expected to be greater for KROSS.

\citet{Johnson2018} also studied a sample of star-forming galaxies from the SAMI Galaxy Survey. They estimated global velocity dispersions for individual galaxies in the range $\sigma_v \sim [20, 60]$ km s$^{-1}$ with one galaxy scattering as high as $\sim$ 90 km s$^{-1}$. Global velocity dispersions as high as 60 km s$^{-1}$ may suggest that they have not fully accounted for beam smearing across all of the galaxies within their sample of galaxies from the SAMI Galaxy Survey. However, given that we have only studied a small sample of galaxies from the SAMI Galaxy Survey, we cannot definitively rule out such high global velocity dispersions.

Another approach to correct for the effects of beam smearing on the observed velocity dispersion is to perform corrections based on the local velocity gradient ($v_\text{grad}$). \citet{Varidel2016} proposed calculating the local velocity gradient using a finite-difference scheme and then performed a regression analysis to estimate the observed velocity dispersion when the local velocity gradient is zero. \citet{Zhou2017} and \citet{Federrath2017} have also used the finite-difference scheme method to remove spaxels where the velocity gradient is much greater than the observed velocity dispersion. We note that \citet{Zhou2017} used this approach to estimate the global H$\alpha$ gas velocity dispersion in the range $\sigma_v \sim [20, 30]$ km $s^{-1}$ with an outlier (GAMA 508421) estimated to be $\sigma_v = 87 \pm 44$ km s$^{-1}$. We note that GAMA 508421 has observed velocity dispersion of $\sim 100$ km s$^{-1}$ in the galaxy centre that that has not been removed. It's possible that this peak is associated with beam smearing. Similarly, \citet{Oliva-Altamirano2018} subtract the local velocity gradient from the observed velocity dispersion in quadrature.

We reproduce these methods on our toy models. First, we revisit the finite-difference scheme and note that the magnitude of the local 1D gradient for a non-boundary spaxel is, 
\begin{equation}
    \bigg| \frac{\partial v}{\partial y} \bigg|_{y_i}
    %v_{\text{grad}, y}(i) 
        \approx \bigg| 
        \frac{ v_{i+1} - v_{i-1} }{ 2 \Delta y }
        \bigg|,
    \label{eq:1DCentralGrad}
\end{equation}
where $i$ is the index and $\Delta y$ is the width of the spaxel in the $y$-direction. The boundary pixels are estimated using the boundary pixel and the adjacent pixel. For a left-sided boundary, the estimated velocity gradient is then, 
\begin{equation}
    \bigg| \frac{\partial v}{\partial y} \bigg|_{y_0}
    %v_{\text{grad}, y}(0) 
        \approx \bigg|
        \frac{ v_1 - v_0 }{ \Delta y }
        \bigg|.
    \label{eq:1DBoundGrad}
\end{equation}
The total absolute magnitude of the velocity gradient is calculated by adding the orthogonal gradients in quadrature,
\begin{equation}
    v_{\text{grad}}(i, j) = \sqrt{
        \bigg| \frac{\partial v}{\partial x} \bigg|_{(i, j)}^2
        + \bigg| \frac{\partial v}{\partial y} \bigg|_{(i, j)}^2
        }.
    \label{eq:2DGrad}
\end{equation}

This expands the previous method to include estimates for the boundary pixels. We also note that within the central pixels the division by 2$\Delta y$ was omitted previously by \citet{Varidel2016}. Strictly speaking, this is incorrect as the gradient will be over-estimated by a factor of 2. Note that the velocity gradient is in units km s$^{-1}$ arcsec$^{-1}$. To make appropriate comparisons between $\sigma_v$ and $v_\text{grad}$, we must convert these to the same units. The most natural scale parameter is the width of the PSF, we choose the $\text{FWHM}_\text{PSF}$ and multiply it by $v_\text{grad}$.

\begin{figure}
    \begin{center}
        \includegraphics[
            width=0.5\textwidth,
            keepaspectratio=true,
            trim=0mm 10mm 0mm 20mm,
            clip=true]{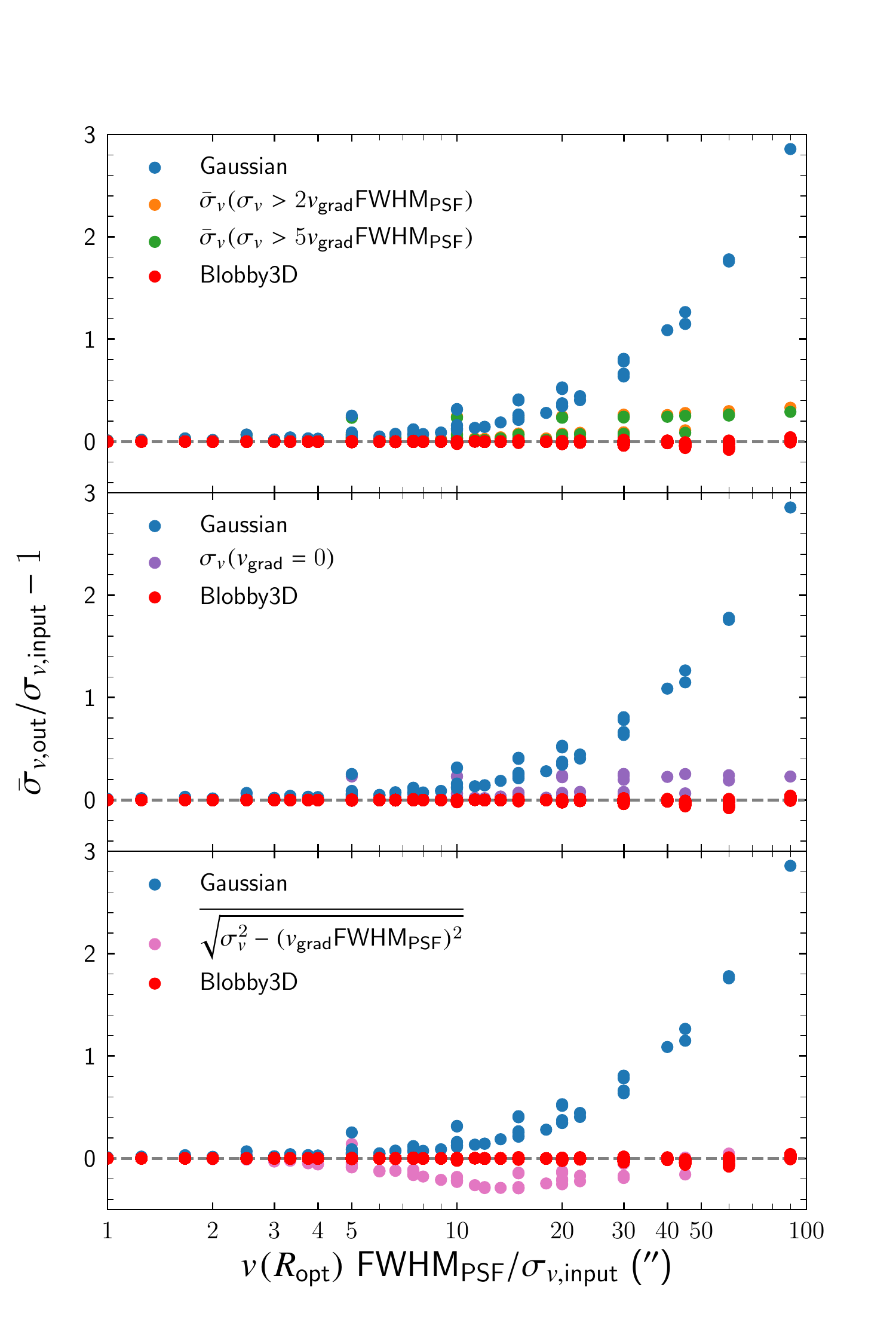}
    \end{center}
    \caption{
    Using heuristic approaches to estimating the mean velocity dispersion for the toy models convolved by a Gaussian PSF using corrections from the observed local velocity gradient ($v_\text{grad}$). Top: estimates in regions where the velocity dispersion is greater than a cutoff value of $\text{FWHM}_\text{PSF} v_\text{grad}(i, j)$. Middle: estimates the velocity dispersion at $v_\text{grad} = 0$ by fitting a cubic to $\sigma_v$ vs $v_\text{grad}$. Bottom: in quadrature subtraction of $v_\text{grad}$ from the observed velocity dispersion. In all cases, these approaches provide significant corrections for the intrinsic mean velocity dispersion compared to the single component Gaussian fits. However, the results from \blobby{} provide the most robust estimates for the intrinsic velocity dispersion.
    }
    \label{fig:VGrandMeanDisp}
\end{figure}

We then repeat the analyses performed previously with the above alterations. We show our results in Fig \ref{fig:VGrandMeanDisp}, including comparison to a single-component Gaussian model per spaxel and our methodology. These methods provide significant corrections from the naive single-component Gaussian fits. However, our method still outperforms these methodologies across our set of toy models.

Over-estimates in regions where beam smearing is high occur for estimates of the mean velocity dispersion where $\sigma_v \gg v_\text{grad} \text{FWHM}_\text{PSF}$. Increasing the cutoff did not result in significantly different estimates of the mean velocity dispersion. Over-estimation is unsurprising as the effect of beam smearing on the observed velocity dispersion occur for several factors of the $\text{FWHM}_\text{PSF}$ where the observed velocity gradient is negligible as seen in Fig. \ref{fig:RadialDispToy}.

For the parametric regression estimates we fit a cubic to $\sigma_v$ vs. $v_\text{grad}$ and then estimated the line at $v_\text{grad} = 0$ km s$^{-1}$ arcsec$^{-1}$. We fit a cubic instead of a first-order line in contrast to \citet{Varidel2016} as there were clear residuals observed by-eye in the linear and quadratic fits to the data. This method suffered from over-estimates of the mean velocity dispersion similar to that observed using the estimates in regions where $\sigma_v \gg v_\text{grad} \text{FWHM}_\text{PSF}$. We suspect this is driven by the observed velocity gradient being shallower than the underlying velocity gradient.

The in quadrature estimates under-estimate the mean velocity dispersions for $5 \lesssim v(R_\text{opt}) \text{FWHM}_\text{PSF} \lesssim 30$. Adjusting a correction parameter $\alpha$ such that the corrections were of the form $\sqrt{ \sigma^2_v - \alpha (v_\text{grad} \text{FWHM}_\text{PSF})^2}$ did not yield significant improvement. We note that \citep{Oliva-Altamirano2018} estimated the local velocity gradient after using \gbkfit{} to estimate the underlying velocity gradient. As such, their estimate for the velocity profile should be less affected by beam smearing, and their velocity gradient will be smooth following a parametric radial profile. They also focused on differences from the mean velocity dispersion, which may not be effected by the precision of the estimate for the global velocity dispersion.

We also note that this is an idealised toy model with negligible noise. In practice, the noise will increase the uncertainties on the local velocity dispersion, which will cause significant deviations in the estimates of the mean velocity dispersion. This could be improved by fitting a velocity profile across the galaxy and using the local velocity gradient derived from that profile similar to \citep{Oliva-Altamirano2018}.

Furthermore, we only applied the velocity gradient approaches to toy models with no gas substructure. As we showed in Fig. \ref{fig:blobby_sample} and \ref{fig:example_sample2}, beam smearing complex gas substructure can have significant effects on the observed gas kinematics. This will effect the estimates for the $v_\text{grad}$, and thus will affect the ability to estimate the underlying velocity dispersion.

These heuristic approaches still provide corrections to the observed velocity dispersion. They are also easy to implement as they use a small number of related parameters (eg.\ velocity gradient, width of the PSF, and distance from the centre of the galaxy). As such, they may be appropriate for particular research purposes. 

As with any heuristic approaches, they often suffer from their simplicity in application. In this case, these methods cannot simultaneously model the beam smearing effect as it acts on the underlying gas and kinematic profiles. They also suffer from not fully taking into account the shape-parameters of the PSF, instead using a single measure of the PSF width such as the FWHM. 3D cube fitting algorithms are the only known approach to the authors that can perform such self-consistent modelling.

\subsubsection{3D cube fitting algorithms}
\label{subsec:3DCubeFit}

There are several 3D cube fitting approaches that have been proposed in the literature. Three of those are publicly available and are specifically designed to work for optical observations. Those are \galpak{} \citep{Bouche2015}, \gbkfit{} \citep{Bekiaris2016}, and \bbarolo{} \citep{DiTeodoro2015}.

As seen in Section \ref{sec:TestModelling}, \bbarolo{} has issues resolving the kinematic profiles in low-resolution observations. This leads to over-estimated velocity dispersion and shallower velocity gradients. Our testing showed no significant difference in the inferred kinematics when running \bbarolo{} with a differing number of rings.

We have no reason to believe that \galpak{} or \gbkfit{} suffer from similar problems. Limitations of \galpak{} and \gbkfit{} are due to the inflexibility of the model parameterisation which will lead to significant residuals for galaxies where complex substructure can be observed. The example galaxies from the SAMI Galaxy Survey seen in Fig. \ref{fig:example_sample1} and \ref{fig:example_sample2} are good examples of such galaxies. An inability to model these complex structures can lead to two potential problems:
\begin{enumerate}
\item The galaxy substructure can be underfit. This can lead to the substructure systematically driving the estimates in indeterminate directions. Underfitting also leads to underestimates of uncertainties \citep{Taranu2017}. 
\item Beam smearing is driven by the smearing of the underlying flux profile. If the underlying flux profile is clumpy it can lead to irregular kinematic profiles as seen in our examples in Fig. \ref{fig:blobby_sample} and \ref{fig:example_sample2}. As such, to get a full understanding of the effects of beam smearing, adequately modelling the gas substructure is important.
\end{enumerate}

We also note that simplifications exist in our current methodology. In particular, assuming the kinematics follow radial profiles is likely to be too simplistic to model a large sample of galaxies. Also, modelling the gas substructure as a hierarchical Gaussian mixture model is also imperfect. We understand that this could lead to similar problems as above. 

The above reasoning led to the introduction of the additional $\sigma_0$ noise term. This term should help account for simple systematic noise between the model and data.

Also the flexibility of using a hierarchical Gaussian mixture model does provide much better fits to the data. To formalise this we performed a Bayesian model comparison between our current methodology with varying number of blobs and setting $N=1$. Setting $N=1$ is similar to a single-component disk model assuming a Gaussian flux profile. In both cases, we calculated the evidence ($Z$) using \dnest. Assuming no prior preference for either model, the odds ratio for our current methodology ($M$) compared to a single component model ($M_0$) is given by $O = p(D|M)/p(D|M_0) = Z/Z_0$. We found $\log(Z/Z_0) = 1.9 \pm 1.2 \times 10 ^{4}$ with $\log(Z/Z_0) > 0$ for all galaxies in our sample from the SAMI Galaxy Survey. Therefore, the variable blob model is preferred compared the single Gaussian component flux model using this measure.

\subsection{Effects of beam smearing on kinematic asymmetries}
\label{sec:KinAsym}

We showed that a toy model with an asymmetric flux distribution, a radial velocity profile, and constant velocity dispersion leads to asymmetries in the velocity dispersion profile once convolved by the PSF (see Section \ref{sec:BlobbyToyModel}). We also saw that modelling of asymmetries in the velocity dispersion profiles of GAMA 30890 can be partially performed by using a flexible gas distribution with radial kinematic profiles plus beam smearing. These observations have implications for the study of asymmetries in observed galaxies.

For example, a popular field of analysis is to estimate the kinematic asymmetries observed in the 2D maps \citep{Shapiro2008}. Analysis of kinematic asymmetries and their drivers have been performed on the SAMI Galaxy Survey previously \citep{Bloom2017a,Bloom2017b,Bloom2018}. In those studies they used \kinemetry{} \citep{Krajnovic2006} to estimate the asymetries in the 2D kinematic maps. \kinemetry{} constructs kinematic maps by interpolating between a series of ellipses. Each ellipse is  decomposed into a Fourier series of the form,
\begin{equation}
    K(a, \psi) = A_0(a)
        + \sum^N_{n=1} ( A_n(a) \sin(n\psi) + B_n(a) \cos(n\psi)),
    \label{eq:kinfourier}
\end{equation}
where $a$ is the semi-major axis length and $\psi$ is the azimuthal angle. This is usually manipulated to the form,
\begin{equation}
    K(a, \psi) = A_0(a) + \sum^N_{n=1} k_n(a) \cos(n (\psi - \phi_n(a)))
    \label{eq:kinfourier2}
\end{equation}
where
\begin{align}
    k_n  = \sqrt{ A_n^2 + B_n^2 } && 
    \text{and} &&
    \phi_n = \arctan \bigg( \frac{A_n}{B_n} \bigg).
    \label{eq:kinkn}
\end{align}
For $n$ is odd the contribution to the 2D map is an even functional contribution. Similarly, for $n$ is even the contribution is an odd functional contribution. The  asymmetric contribution to a kinematic moment per spaxel is typically calculated using a ratio of the sum of $k_{n, \text{mom}}$ for n > 1 compared to the first-order velocity moment $k_{1, v}$. In previous works on data from the SAMI Galaxy Survey, the following has been used,
\begin{align}
    v_{\text{asym}} = \frac{k_{3, v} + k_{5, v}}{2 k_{1, v}} &&
    \text{and} &&
    \sigma_{v, \text{asym}} = \frac
        {k_{2, \sigma_v} + k_{4, v}}
        {2 k_{1, v}}.
    \label{eq:bloomvasym}
\end{align}
The odd moments were ignored for $v_{\text{asym}}$ and the even moments were ignored for $\sigma_\text{asym}$ as they were estimated to be negligible.

Analysing a sample of 360 galaxies \citet{Bloom2017a} estimated the mean asymmetry across the FoV to be $\bar{v}_{\text{asym}} = 0.044^{+0.044}_{-0.017}$ and $\bar{\sigma}_{v, \text{asym}} = 0.10^{+0.17}_{-0.04}$. This suggests greater asymmetries in the velocity dispersion compared to the velocity maps. However, the effect of beam smearing on the kinematic asymmetries has not been investigated. 

Expanding our method to account for asymmetries in the  velocity and velocity dispersion profiles would allow for simultaneous fitting of the kinematic asymmetries while taking into account the effects of beam smearing. This could be achieved by adopting the Fourier series decomposition of the moments similar to \kinemetry. A natural way to do so would be to parameterise $k_n$ and $\phi_n$ as radial functions across the disk.

We also note that \citet{Bloom2017a} assigned $23\pm7\%$ of 360 galaxies from the SAMI Galaxy Survey as perturbed. In their analysis, they assigned galaxies to be perturbed when $\bar{v}_{\text{asym}} > 0.065$. Thus, accounting for asymmetries is an important factor in accurately modelling a larger sample of galaxies at similar resolutions to the SAMI Galaxy Survey.

\subsection{Implications for the study of gas turbulence within galaxies}
\label{subsec:imp}

Observations have established that galaxies at  $z > 1$ exhibit higher velocity dispersion as well as clumpier gas and velocity dispersion profiles \citep{Genzel2011,Wisnioski2011} compared to local galaxies. As the PSF width relative to the observed galaxy size is greater at higher redshift, the effects of beam smearing will typically be greater. As such, it is possible to mistakenly draw correlations across epochs if the effects of beam smearing on the gas velocity dispersion have not been corrected.

One relevant claim has been that star-formation feedback processes play an important role as a driver of gas turbulence across epochs \citep{Green2010,Green2014}. In contrast, there have been several studies of the localised star-formation rate and gas turbulence in nearby galaxies which have not found a significant correlation \citep{Varidel2016,Zhou2017}. Another recent claim has been that gas turbulence may be driven by the interaction between clumps and the interstellar medium \citep{Oliva-Altamirano2018}. Inferring these relationships requires an ability to accurately determine the intrinsic gas distribution and kinematics. In such studies, our approach would provide a measure for the intrinsic velocity dispersion while taking into account the potentially complex gas distribution.

In particular, inferring relationships between gas clumps and the local kinematics should be much easier in our approach. For example, the study of the residuals in the velocity dispersion map could indicate clear peaks in the velocity dispersion that are correlated with the intrinsic gas distribution. A more natural way within the Bayesian framework, would be to parameterise the velocity dispersion as a function of the gas flux. The simplest approach would be to assume a velocity dispersion component of the form $\log(\sigma_v) \propto F(x, y)$, where the proportionality constant would be a free parameter.

\subsection{Potential applications for the study of gas outflows}
\label{subsec:outflows}

Gas outflows play an important role as a star-formation feedback mechanism  \citep{Elmegreen2009,Federrath2017IAUS}. As such, the identification of gas outflows in star-forming galaxies has received considerable attention \citep[eg.][]{Ho2014}.

A difficulty in studying gas outflows is to distinguish between the gas rotation, gas outflows, and contributions of beam smearing on the observed emission line profiles. We suggest that applications of forward fitting modelling approaches, such as \blobby{}, are ideal to study these galaxies as the rotation and beam smearing contributions can be taken into account simultaneously. 

In ideal circumstances, it will be possible to identify outflows as residuals from the 3D model. However, an ideal extension to \blobby{} for the study of gas outflows, would be to construct a parametric model for the gas outflows. This parameterisation would need to be carefully constructed as winds do not follow the rotational gas kinematics. As such, gas outflows would introduce asymmetries in the emission line profiles with different geometries to the galaxy plane. 

This may require the introduction of higher-order moments for the emission line profiles. Functional forms for the emission lines that could be used are skewed Gaussian or Hermite-Gaussian profiles. An alternative approach would be to add a secondary gas velocity and velocity dispersion profile which has characteristics that represent an outflow. A simplistic model would likely require a parameterisation for the gas component moving radially outwards in a cone-like shape with a given velocity and velocity dispersion profile.

\subsection{A note on run time}
\label{subsec:RunTime}

Other 3D fitting algorithms take $\mathcal{O}$(seconds - minutes) to run a typical SAMI Galaxy Survey sized cube cut around the H$\alpha$ emission line. The current \textsc{C++} implementation of \blobby{} took the equivalent of $\sim$450 Central Processing Unit (CPU) hours for a single galaxy within our SAMI Galaxy Survey sample. Wall time was reduced significantly by running \dnest{} in multi-threaded mode.

The run time is a function of the complexity of the gas substructure, the signal-to-noise, and the number of samples saved. The run time is a considerable disadvantage for researchers that have very large data sets or are low in computing resources. We have been able to work around this issue by using the Artemis cluster provided by The University of Sydney HPC Service. This gave us access to a large number of cores, such that we could run our methodology for several galaxies simultaneously.

The bottleneck is primarily driven by the number of blobs required to construct the flux profile. Thus decreasing the maximum number of blobs ($N_\text{max}$) will decrease the run time significantly. Of course, this will lead to posterior distributions for $N$ being abruptly cut-off at $N_\text{max}$ for some galaxies. We could also implement non-uniform priors for the number of blobs. Similarly, some researchers may find that decomposing the gas distribution into a fixed number of blobs will be adequate to model the gas substructure. In these cases, the prior space will be significantly decreased, and thus will result in significantly faster convergence. We have not explored these possibilities in this work, but it may be important as we scale the methodology to larger samples.

Another approach would be to use an optimisation routine compared to a sampling algorithm. In this case, the user would only get an optimised point estimate, but such algorithms are typically much quicker. We note that there is an ability to optimise using  \dnest{}. We have avoided optimisation techniques as we prefer to perform the full inference in order to estimate uncertainties.

Despite the improvements in speed that could be made, we still expect that our method will be significantly slower than other similar 3D fitting algorithms. However, the time restrictions implicit in our method are offset by the improvements in modelling the complex gas substructure that is apparent in typical IFS observations. Furthermore, due to the effects on kinematics that were discussed in Section \ref{sec:KinAsym}, we suggest that researchers should consider using such flexible modelling approaches for the gas substructure in order to accurately infer the intrinsic gas kinematics in their observations.

\section{Conclusions}
\label{sec:Conclusions}

Beam smearing occurs due to the flux profile being spread-out across the FoV by the seeing. For rotating disks this has significant effects on the observed kinematics. It has been well known that the observed LoS velocity profiles are typically flattened and the LoS velocity dispersion is increased when assuming single flux component galaxy models \citep{Davies2011}.

However, the observed gas distribution often exhibits complex structure including clumps, rings, or spiral arms. Considering this fact, we developed a methodology referred to as \blobby. \blobby{} can model complex gas substructure by using a hierarchical Gaussian mixture model. The kinematics are modelled assuming radial profiles. We take into account the effect of beam smearing by convolving the model by the seeing per spectral slice before comparing it to the data.

\blobby{} was applied to a sample of 20 star-forming galaxies from the SAMI Galaxy Survey. We estimated the global gas velocity dispersions for all galaxies in the range $\bar\sigma_v \sim $[7, 30] km s$^{-1}$. This is in comparison to estimates using a single Gaussian component per spaxel that were in the range $\bar\sigma_v \sim$[10, 45] km s$^{-1}$. The relative corrections per galaxy were $\Delta \sigma_v / \sigma_v = -0.29 \pm 0.18$. This has implications for galaxies observed at $z > 1$ that have observed gas velocity dispersions typically much greater than nearby galaxies.

We also show that resolving the gas substructure is important as the gas substructure can lead to asymmetries in the kinematic profiles. A toy model was constructed with asymmetric gas substructure with radial kinematic profiles plus beam smearing to show that asymmetric substructure was observable in the observed gas velocity dispersion. We also found that asymmetries in the velocity dispersion maps for GAMA 30890 can be partially recovered using our methodology, that only introduces asymmetries in the gas distribution. This implies that studies of asymmetries within galaxies should consider the effects of beam smearing on their results.

To accurately infer the intrinsic gas kinematics both the gas flux and kinematic profiles plus beam smearing should be considered. With this in mind, methods such as \blobby{}, that are capable of performing such inferences should be an important step in analysing the kinematics for IFS observations of gas disks.

\section*{Acknowledgements}

The SAMI Galaxy Survey is based on observations made at the Anglo-Australian Telescope. The Sydney-AAO Multi-object Integral field spectrograph (SAMI) was developed jointly by the University of Sydney and the Australian Astronomical Observatory. The SAMI input catalogue is based on data taken from the Sloan Digital Sky Survey, the GAMA Survey and the VST ATLAS Survey. The SAMI Galaxy Survey is supported by the Australian Research Council Centre of Excellence for All Sky Astrophysics in 3 Dimensions (ASTRO 3D), through project number CE170100013, the Australian Research Council Centre of Excellence for All-sky Astrophysics (CAASTRO), through project number CE110001020, and other participating institutions. The SAMI Galaxy Survey website is \url{http://sami-survey.org/}.

The authors acknowledge the University of Sydney HPC service at The University of Sydney for providing HPC and database resources that have contributed to the research results reported within this paper. 
URL: \url{http://sydney.edu.au/research_support/}

BJB acknowledges funding from New Zealand taxpayers via the Marsden Fund of the Royal Society of New Zealand. JBH is supported by an ARC Laureate Fellowship that funds JvdS and an ARC Federation Fellowship that funded the SAMI prototype. EDT acknowledges the support of the Australian Research Council (ARC) through grant DP160100723. JJB acknowledges support of an Australian Research Council Future Fellowship (FT180100231). CF acknowledges funding provided by the Australian Research Council (Discovery Projects DP170100603 and Future Fellowship FT180100495), and the Australia-Germany Joint Research Cooperation Scheme (UA-DAAD). BG is the recipient of an Australian Research Council Future Fellowship (FT140101202). Support for AMM is provided by NASA through Hubble Fellowship grant \#HST-HF2-51377 awarded by the Space Telescope Science Institute, which is operated by the Association of Universities for Research in Astronomy, Inc., for NASA, under contract NAS5-26555. MSO acknowledges the funding support from the Australian Research Council through a Future Fellowship (FT140100255). NS acknowledges support of a University of Sydney Postdoctoral Research Fellowship.

%%%%%%%%%%%%%%%%%%%%%%%%%%%%%%%%%%%%%%%%%%%%%%%%%%



\bibliographystyle{mnras}
\bibliography{Blobby3D.bib} % if your bibtex file is called example.bib

%\begin{thebibliography}{99}
%\bibitem[\protect\citeauthoryear{Bouch\'e et al.}{2015}]{Bouche2015}


%\end{thebibliography}

%%%%%%%%%%%%%%%%%%%%%%%%%%%%%%%%%%%%%%%%%%%%%%%%%%

%%%%%%%%%%%%%%%%% APPENDICES %%%%%%%%%%%%%%%%%%%%%

%\appendix


% If you want to present additional material which would interrupt the flow of the main paper,
% it can be placed in an Appendix which appears after the list of references.

%%%%%%%%%%%%%%%%%%%%%%%%%%%%%%%%%%%%%%%%%%%%%%%%%%

% Don't change these lines
\bsp	% typesetting comment
\label{lastpage}
\end{document}